\documentclass{emulateapj}

\usepackage{txfonts}
\usepackage{float}
\usepackage[usenames,dvipsnames]{color}
\usepackage{ulem}

\slugcomment{}

\shorttitle{Dynamical Evolution of Viscous Disks around Be Stars}
\shortauthors{Haubois et al.}

\begin{document}


\title{Dynamical Evolution of Viscous Disks around Be Stars. II : polarimetry.}


\author{X. Haubois\altaffilmark{1,2,3}}
\author{B. C. Mota\altaffilmark{2}}
\author{A. C. Carciofi\altaffilmark{2}}
\altaffiltext{1}{LESIA, Observatoire de Paris, CNRS UMR 8109, UPMC, Universit\'e Paris Diderot, 5 place Jules Janssen, F-92195 Meudon, France, \email{xavier.haubois@obspm.fr}}
\altaffiltext{2}{Instituto de Astronomia, Geof\'{i}sica e Ci\^{e}ncias Atmosf\'{e}ricas, Universidade de S\~{a}o Paulo, Rua do Mat\~{a}o 1226, Cidade Universit\'{a}ria, S\~{a}o Paulo, SP 05508-090, Brazil}
\altaffiltext{3}{Sydney Institute for Astronomy, School of Physics, University of Sydney, NSW 2006, Australia}
\author{Z. H. Draper\altaffilmark{4,5}}
\altaffiltext{4}{Department of Physics and Astronomy, University of Victoria, 3800 Finnerty Rd, Victoria, BC V8P 5C2 Canada}
\altaffiltext{5}{Herzberg Institute of Astrophysics, National Research Council of Canada, Victoria, BC V9E 2E7 Canada}
\author{J. P. Wisniewski\altaffilmark{6}}
\altaffiltext{6}{H. L. Dodge Department of Physics and Astronomy, University of Oklahoma, 440 West Brooks St Norman, OK 73019, USA}
\author{D. Bednarski\altaffilmark{2}}
\author{Th. Rivinius\altaffilmark{7}}
\altaffiltext{7}{European Organisation for Astronomical Research in the Southern Hemisphere, Casilla 19001, Santiago 19, Chile}

%\author{J. E. Bjorkman}
%\affil{University of Toledo, Department of Physics \& Astronomy, MS111 2801 W. Bancroft Street Toledo, OH 43606, USA}
%\and
%\author{A.T. Okazaki}
%\affil{Faculty of Engineering, Hokkai-Gakuen University, Toyohira-ku, Sapporo 062-8605, Japan}


%\altaffiltext{1}{}
%\altaffiltext{2}{}
%\altaffiltext{3}{}
%\altaffiltext{4}{}


\begin{abstract}
Be stars exhibit variability for a great number of observables. Putting the pieces together of the disk dynamics is not an easy task and requires arduous modeling before achieving a good fit of the observational data. In order to guide the modeling process and make it more efficient, it is very instructive to investigate reference dynamical cases. This paper focuses on continuum polarimetric quantities and is the second of a series that aims to demonstrate the capacity of deriving the dynamical history and fundamental parameters of a classical Be star through the follow-up of various observables. After a detailed study of the different opacities at play in the formation of polarized spectra, we investigate predictions of polarimetric observables in the continuum for different dynamical scenarios. Our models are based on a coupling of a hydrodynamic viscous decretion simulations in a disk and a 3-D non-LTE radiative transfer code. Introducing the polarization color diagram (PCD), we show that certain combinations of polarimetric observables exhibit features that are characteristic of a mass loss history. This diagram also enables to estimate fundamental parameters such as the inclination angle, the disk density scale and the $\alpha$ viscous diffusion parameter. We present the PCD as a powerful diagnosis tool to track the dynamical phases of a Be star such as disk building-up, dissipation, periodic and episodic outbursts. Finally we confront our models with observations of 4 Be stars that exhibited long-term polarimetric activity.

%Which band, niveau de polarization attendu, quels diagrammes sont caracteristiques de quels scenarios, alpha ?? 

%   comparaisons aux vrais donnees : ca se rapporte a des combinaisons de cas classiques ?

\end{abstract}

\keywords{circumstellar matter --- radiative transfer -- stars: emission-line, Be  --- stars: individual ($\pi$~Aquarii, 60~Cygni, $\delta$ Scorpii and $\psi$~Persei)  --- techniques: polarimetric}

 \section{Introduction}

Be stars are non-supergiant, early-type stars with a circumstellar (CS) disk that is created from matter ejected from the star. Recent observational facts brought by spectro-interferometry and spectro-astrometry \citep[e.g.][]{2012A&A...538A.110M,2012MNRAS.423L..11W} supports the fact that so far all studied Be star disks rotate in a Keplerian fashion. This characteristic, together with other observational signatures of the disk outlined in \cite{2011IAUS..272..325C}, are properties that only the viscous decretion disk  (VDD) model can reproduce. This model, first suggested by \cite{1991MNRAS.250..432L} and further developed by \citet{1997LNP...497..239B}, \citet{1999A&A...348..512P}, \citet{2001PASJ...53..119O}, \citet{2005ASPC..337...75B} and \citet{2008MNRAS.386.1922J}, among others, uses the angular momentum transport by turbulent viscosity to lift material into higher orbits, thereby causing the disk to grow in size. This model has already been successfully applied to systems showing stable continuum emission: e.g. $\zeta$ Tauri \citep{car09}, $\chi$\,Oph \citep{2008ApJ...689..461T} and $\beta$\,CMi \citep{2012MNRAS.423L..11W} and systems exhibiting a more variable photometric activity \citep[28\,CMa,][]{2012ApJ...744L..15C}.  In a recent review paper, \cite{rev2013} discuss in detail the observational and theoretical evidences in support of the VDD scenario for Be stars.

Polarization is a powerful tool to study the geometry of the disk (opening angle, flaring) of Be stars without angularly resolving it. Polarized flux originates from electron scattering off the disk, and is affected by both pre- and post-scattering absorption by \ion{H}{1} atoms \citep{wood96b,2013ApJ...765...17H}. Since \ion{H}{1} opacity depends on the physical state of the gas, studying polarimetric observables at different wavelengths allows one to probe different regions of the disk. In the literature, the polarimetric technique has an established history of providing a unique diagnosis in identifying and studying the detailed CS environments of Be stars \citep[][]{1997ApJ...477..926W,2007ApJ...671L..49C,car09,2010ApJ...709.1306W,2011ApJ...728L..40D}.

Haubois et al. (2012), hereafter Paper I, studied the temporal variability of Be disks, based on {\sc singlebe} VDD hydrodynamics simulations  \citep{2007ASPC..361..230O}. {\sc singlebe} solves the 1-D surface density evolution equation for a viscous isothermal Keplerian decretion disk. The effects of variable mass injection rates on the disk structure, and their corresponding effect on the photometry, were studied at different wavelengths and compared to observations. More specifically, we first studied the different timescales that characterize the evolution of the disk surface density and how this surface density responds to changes in the mass injection rates. These surface density profiles were used as inputs to the three-dimensional non-LTE Monte Carlo radiative transfer code {\sc hdust}  \citep{car06,2008ApJ...684.1374C} that allowed the calculation of photometric observables at various wavelengths. The characteristic shapes of these lightcurves agree qualitatively well with observations, which provide strong circumstantial evidence that viscosity is indeed the mechanism that redistributes matter along the CS disk. The first successful confrontation of theoretical VDD lightcurves with observations was done by \cite{2012ApJ...744L..15C} for the Be star 28\,CMa.
To summarize, Paper I  provides the reader with a description of the photometric variability from a Be star in the framework of the VDD model. With the present paper, we aim at  exploring the variability of the  \emph{continuum} polarimetric features in the same manner.

%We then computed lightcurves for Be stars subject to different mass loss rate histories and studied how the lightcurves depended on the other system parameters. 
 
In \S÷\ref{origin}, we describe the polarigenic mechanisms operating in Be disks. Then we present the dynamical models we investigated and their signatures on common continuum polarimetric observables. We also show the diagnosis potential of a series of diagrams that we named polarization color diagrams (PCDs) that represent a powerful tool to follow the mass injection history in Be stars (\S÷\ref{prediction}). Finally, a discussion and a comparison to observed data of these synthetic observables are presented in \S÷\ref{comp} before concluding.

%mass injection = (mass ejected from the central star and injected into the disk)

\section{Polarization in Be star disks}
\label{origin}
Continuum spectropolarimetric observations of Be stars usually reveal a sawtooth pattern that displays abrupt changes of the polarization close to the \ion{H}{1} ionization thresholds \citep[see, e.g.,][for examples of observed polarization spectra]{1997ApJ...479..477Q}. It is useful to review the origin of this pattern to understand precisely the physical processes that control the shape of the polarized spectrum. In this Section, we therefore adopt an analytical model to describe a viscous decretion disk surrounding a rotationally deformed and gravity darkened star, see Table~\ref{tab_param} for the adopted stellar parameters.

% \begin{deluxetable}{lcccccc}
%% \tableline\tableline
%\tablecolumns{7}
%\tablecaption{Stellar main parameters used in the simulations. \label{tab_param}}
%\tablehead{\colhead{Parameter} & \colhead{B0} & \colhead{B1}& \colhead{B2}& \colhead{B3}& \colhead{B4}& \colhead{B5}}
%\startdata
%Mass ($M_{\odot}$) &   && 9.0    &  \\
%Polar radius ($R_{\odot}$) & &&5.7  & & & \\
%Equatorial radius  $R_{\star}$ ($R_{\odot}$) & &&  6.5  & & &  \\ %delta sco
%Rotation speed (km/s) & &&273  & & & \\
%$V_{K}$\tablenotemark{1} (km/s)&   & &514 & & & \\
% $V_{c}$ \tablenotemark{2} (km/s) &    & &448 & & & \\
%$\Omega$ / $\Omega_{c}$ &  &&0.8 & & & \\ %273*(1.5/1.14)/448.
%Oblateness &&& 1.14 & & &  \\
%Polar temperature (K) &&& 22000 & & & \\ % sigma Ori ?
%%Ratio between the polar and equatorial temperature & 1.16 \\ 
%Luminosity ($L_{\odot}$) & &&5980  & &  &\\ 
%%\tableline
%\enddata
%\tablenotetext{1}{Keplerian speed at equator}
%\tablenotetext{2}{Breakup speed, $V_{c}=\sqrt{\frac{2}{3}GM/R_{pol}}$}
%\end{deluxetable}

\begin{deluxetable}{lcccccc}
% \tableline\tableline
\tablecolumns{7}
\tablecaption{Stellar main parameters used in the simulations. \label{tab_param}}
\tablehead{
\colhead{Parameter}                          & \colhead{B0}  &  \colhead{B1}  &  \colhead{B2}  & \colhead{B3}  &  \colhead{B4}  &  \colhead{B5}} \startdata
Mass ($M_{\odot}$)                           & $14.6$        & $11.0$         & $8.6$          & $6.1$         & $5.1$ 	        & $4.4$           \\ 
Polar radius $R_{pole}$($R_{\odot}$)                   & $12.8$        & $9.6$          & $7.5$          & $5.4$         & $4.5$          & $3.8$            \\
Equatorial radius  $R_{\star}$ ($R_{\odot}$) & $5.8$         & $4.9$          & $4.3$          & $3.6$         & $3.3$ 	        & $3.0$             \\
W \tablenotemark{1}                          & $0.53$        & $0.53$         & $0.53$         & $0.53$        & $0.53$  	& $0.53$             \\ 
$v_{\rm rot}$ (km/s)            	     & $344$         & $325$          & $308$          & $283$         & $272$          & $261$               \\
$v_{\rm orb}$ (km/s) 	                     & $648$         & $613$          & $580$          & $534$         & $512$      	& $492$	               \\
%$v_{\rm crit}$\tablenotemark{2} (km/s) 	     & $565$         & $534$          & $506$          & $466$         & $447$      	& $429$	                \\
$\Omega$ / $\Omega_{\rm crit}$                  & $0.8$         & $0.8$          & $0.8$          & $0.8$         & $0.8$          & $0.8$                  \\
Oblateness                                   & $1.14$        & $1.14$         & $1.14$         & $1.14$        & $1.14$         & $1.14$                  \\
Polar temperature (K)                        & $29\,900$     & $26\,200$      & $23\,100$      & $19\,100$     & $17\,200$  	& $15\,500$                \\ 
Luminosity ($L_{\odot}$)                     & $24\,200$     & $10\,200$      & $4\,400$       & $1\,500$      & $830$ 	        & $467$                     
\enddata
\tablenotetext{1}{$W = v_{\rm rot}/ v_{\rm orb}$, see Sect. 2.3.1 of \cite{rev2013}}
%\tablenotetext{2}{Breakup speed, $v_{crit}=\sqrt{\frac{2}{3}G\,M/R_{pole}}$}
\end{deluxetable}

For the disk, we adopted the VDD model in its simplest form: after a sufficiently long and stable period of mass decretion, a viscous disk assumes a power-law density profile given by $\rho = \rho_0\,(R_\star/r)^{3.5}$  \citep{2005ASPC..337...75B}, where $\rho_0$ is the density at the base of the disk. In this study and in the rest of the paper, we keep the disk outer radius fixed at 20 $R_\star$. Thus, in this Section we explore the effects of two model parameters in the polarized spectrum: the spectral type of the star and the disk density scale.

%We will study the polarized spectrum for a series of models that are representative of typical Be stars. Our model is composed of a central star surrounded by a viscous gaseous disk. For the central star we chose parameters that describe the spectral types B0 - B5 and luminosity class V. Each star is described by three physical parameters (mass, polar radius and polar effective temperature) and an angular rotation rate, for which we adopt 80\% of the critical angular speed \citep{2003PASP..115.1153P}.

 %We adopted the case of an isothermal viscous diffusion in the steady-state regime. This state is reached after a sufficiently long and stable decretion period (Paper I) and the radial disk surface density $\Sigma$ then takes the following simple form: $\Sigma ( r)$  $\propto r^{-2}$ \citep{2005ASPC..337...75B}.  

%Parameter Symbol Ref. Case
%Spectral type - B1V
%Polar radius Rpole 4.9 R
%Pole temperature Tpole 27440 K
%Luminosity L? 10160 L
%Critical velocity vcrit 534.4 km/s
%Rotation rate  om/omeg crit 0.8
%Oblateness Req/Rpole 1.14
%Gravity darkening Tpole/Teq 1.16
%Disk radius Rdisk 10 R?
%Disk density n 1013 cm?3

\begin{figure}[h!]
% \vspace*{-2.0 cm}
\begin{center}
 \includegraphics[width=3.2in]{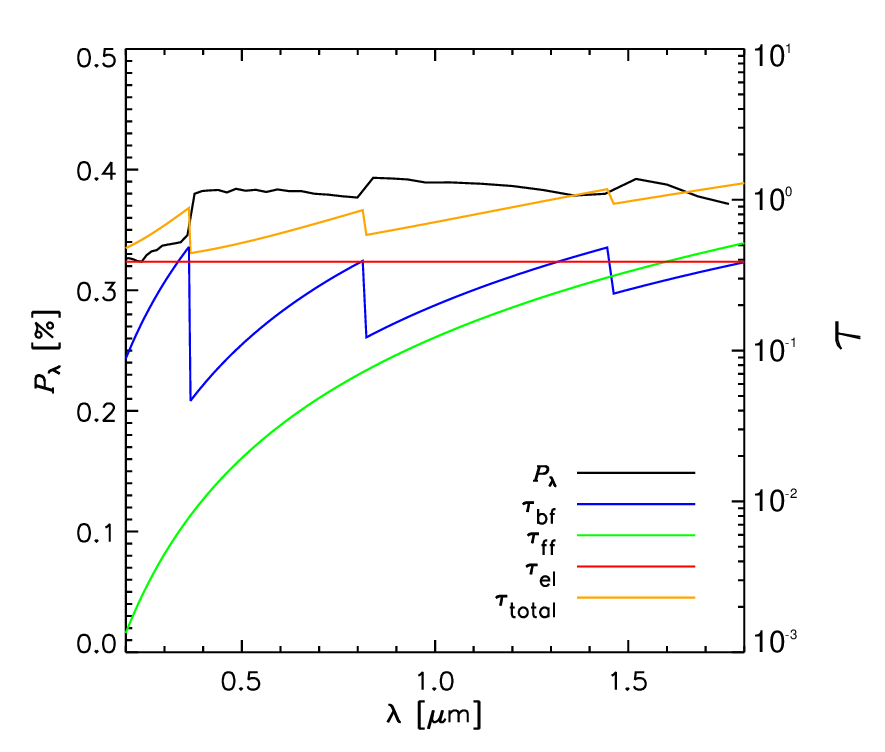} 
 \includegraphics[width=3.2in]{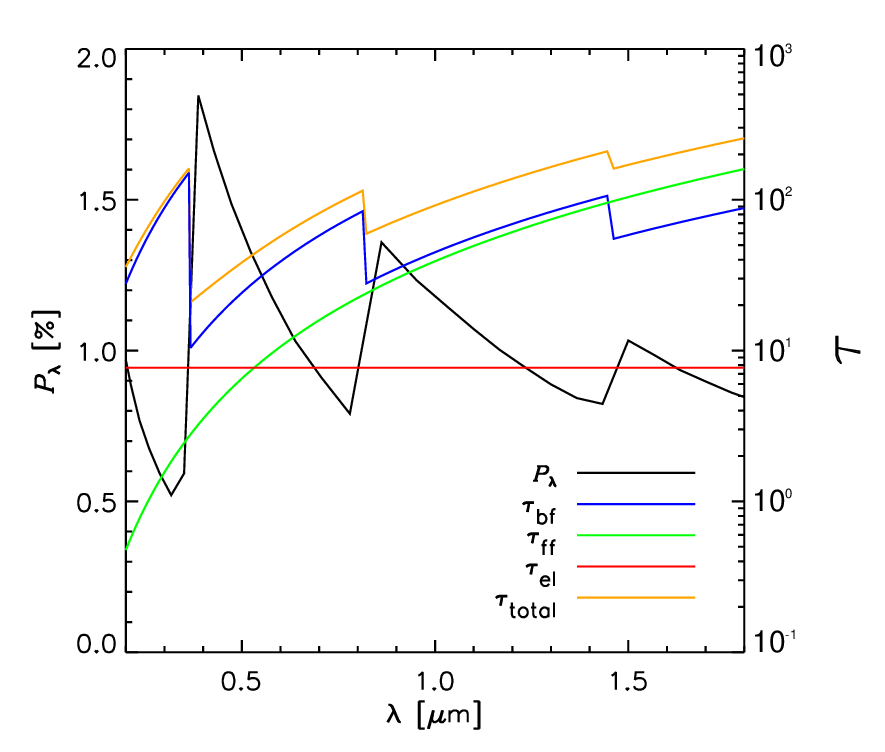}  
% \includegraphics[width=3.2in]{modelo4.eps}  
 %\caption{Optical depths (bound-free $\tau_{bf}$, free-free $\tau_{ff}$ and electronic scattering $\tau_{el}$) and polarized spectrum $P_{\lambda}$ as functions of the wavelength for two densities at the base of the disk around a B2 star (upper panel:  $1.5\times10^{-11}g.cm^{-3}$,  lower panel: $3\times10^{-10}g.cm^{-3}$ ). The inclination angle is 70$\degr$.}
\caption{Polarized spectrum and radial optical depth contributions along the midplane. The total optical depth is the sum of the optical depth for each continuum opacity source, as indicated. Upper panel: $4.2\times10^{-12}\, \rm g \,cm^{-3}$; lower panel: $8.4\times10^{-11}\,\rm g\,cm^{-3}$. The inclination angle is 70$\degr$ and the spectral type is B2.}
   \label{tau_lambda}
\end{center}
\end{figure}

Figure~\ref{tau_lambda} shows two theoretical polarized spectra, at two different base densities and for an inclination angle of 70 $\degr$. While the low density model has a nearly flat spectrum in the optical and near IR, the high density model displays a steep spectrum with marked changes in the \ion{H}{1} ionization thresholds. 
This behavior can be understood in terms of the relative contribution to the total opacity of each opacity source. Figure~\ref{tau_lambda} shows the total optical depth of the disk, measured in the radial direction along the midplane (orange lines). The contribution of the absorptive (free-free and bound-free) and scattering (Thomson) opacities are also shown.

%Let's study two different cases of density at the base of a B2e star, one with a high density of 10$^{14}$ particles per cm$^{3}$ and the other one with 10$^{13}$ particles per cm$^{3}$. 

%With the exception of short part located at $\lambda < 91.2$ nm (below the Lyman discontinuity) where the \ion{H}{1} absorption overtakes the electronic scattering, the optical depth spectrum is basically flat at those wavelengths. (for 1 e12 p/cm3 figure)
At low density, the electron scattering opacity, which is wavelength independent, is responsible for most of the total opacity. The resulting polarized spectrum is consequently nearly flat. However, changes in the polarization level does occur close to \ion{H}{1} ionization thresholds, more importantly at the Balmer discontinuity (0.365\,$\mu\rm m$). What causes the decrease in the polarization redward of the discontinuities is the increase in the \ion{H}{1} opacity. This effect is thoroughly discussed in \citet{wood96b}, and can be understood in terms of pre-scattering (and to a lesser degree, post-scattering) absorption of starlight, that decreases the polarized flux and hence the polarization level, creating an anti-correlated aspect of the optical depth and polarization curves. If the density of the disk increases, the electron opacity will increase more or less linearly, as it is proportional to the number density of free electrons ($\rho$-diagnostics). However, the bound-free and free-free opacities, being roughly proportional to the \emph{square} of the density \citep[e.g., equation~\ref{bfsct} and Eq.~30 of][]{1994ApJ...436..818B}, will increase much faster than the electron opacity ($\rho^2$-diagnostics) so that, at high disk densities (lower panel of Fig.~\ref{tau_lambda}), the bound-free opacity dominates the total opacity. As a result, the polarized spectrum, besides displaying much more pronounced changes in the \ion{H}{1} discontinuities, has a quite steep slope that results from the spectral dependency of the bound-free opacity. 

 \begin{figure}[t]
% \vspace*{-2.0 cm}
\begin{center}
\includegraphics[width=3.5in]{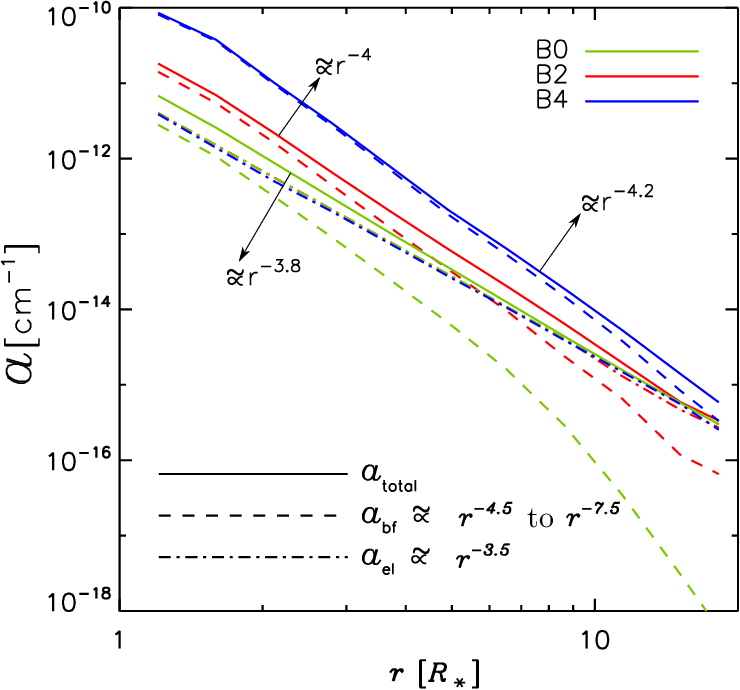} 
 \caption{Absorption coefficient redwards of the Balmer discontinuity as a function of the distance to the stellar surface. %Upper panel: the calculations assume a rotating B2 star surrounded by a viscous decretion disk with a base density of $\rho_{0}$ = $8.4\times10^{-12} \rm g \,cm^{-3}$ (red curves) and $\rho_{0}$ = $4.2\times10^{-11} \rm g \,cm^{-3}$ (blue curves). 
%Lower panel: 
The calculations assume a base density of $\rho_{0}$ = $8.4\times10^{-12} \rm g \,cm^{-3}$ and three spectral types: B0 (green curve), B2 (red curve) and B4 (blue curve). The free-free absorption coefficient is not shown because for short wavelengths it is much smaller than the other opacities. The electron scattering opacities are very similar for the three spectral types so that their curves overlap. Power-law indexes were estimated for the individual (in the legend) and total (on the graph) absorption coefficients.}
\label{loci}
\end{center}
\end{figure}
%source: V_rot =  310.3 km/s
% mod02_B2_4.0: criticial velocity, V_crit = 585.2 km/s

%
% \begin{figure}[h!]
%% \vspace*{-2.0 cm}
%\begin{center}
%\includegraphics[width=3.5in]{tauxden_1bis.eps} 
% \caption{XXXXXXX Absorption coefficient redwards of the Balmer discontinuity as a function of the distance to the star surface. %Upper panel: the calculations assume a rotating B2 star surrounded by a viscous decretion disk with a base density of $\rho_{0}$ = $8.4\times10^{-12} \rm g \,cm^{-3}$ (red curves) and $\rho_{0}$ = $4.2\times10^{-11} \rm g \,cm^{-3}$ (blue curves). 
%%Lower panel: 
%The calculations assume a base density of $\rho_{0}$ = $8.4\times10^{-12} \rm g \,cm^{-3}$ and two spectral type: B0 (black curve) and B3 (blue curve). The free-free absorption coefficient is not shown because for short wavelengths it is much smaller than the other opacities. The electron scattering opacities are very similar for the two spectral types so that their curves overlap.
%}
%\label{loci_dens}
%\end{center}
%\end{figure}

\begin{figure}[b]
% \vspace*{-2.0 cm}
\begin{center}
 \includegraphics[width=3.5in]{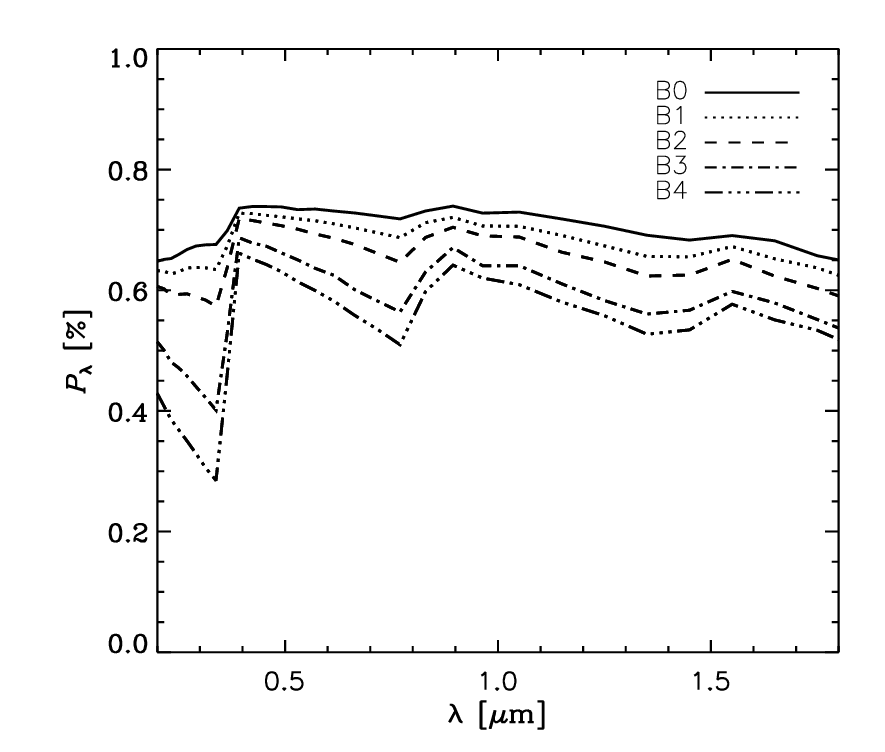} 
 \caption{Polarized spectra for 5 spectral types: B0 (top curve) to B4 (bottom curve). The base density of the disk is  $8.4\times10^{-12} \rm g \,cm^{-3}$. Inclination angle is $70\degr$. }
   \label{polxst}
\end{center}
\end{figure}

In addition to the total radial optical depth, another important quantity in shaping the polarized spectrum is the radial dependence of the opacities or equivalently absorption coefficients. These are shown in Figs.~\ref{loci} and \ref{loci_dens} for a wavelength redwards of the Balmer discontinuity. The effects of the spectral type on the absorption coefficients are shown in Fig.~\ref{loci}. The slope of the electron absorption coefficient curve, $a_{\rm el}$, follows roughly the slope of the density ($\propto r^{-3.5}$), because, for the models shown, \ion{H}{1} is more than 98\% ionized everywhere in the disk (in other words, the ionization fractions are close to unity and, therefore, the electron number density is roughly proportional to the total density). However, the bound-free absorption coefficient, $a_{\rm bf}$, falls much faster than the electron absorption coefficient. For the B4 model, for instance, $a_{\rm bf} \propto r^{-4.5}$, and no significant difference in slope was found for other spectral types (Fig.~\ref{loci}). The value of the radial slope of the bound-free opacity is controlled by the \ion{H}{1} level populations, as shown in Appendix~\ref{appendix}. The actual values of the opacities differ quite markedly for different spectral types: the later the spectral type the larger the opacity. This is explained by the changing ratio between the ionizing UV luminosity vs. the total luminosity. Since the electron opacity is essentially the same for the models shown in Fig.~\ref{loci}, the large differences in the bound-free opacity implies that the dominant opacity source is different for each spectral type. So, for a B0 star, electron scattering dominates at $0.3647\,\rm \mu m$ everywhere in the disk, and the opposite is true for the B4 star. An intermediate behavior is seen for the B2 star model: while the bound-free absorption coefficient dominates in the inner disk (up to $\approx 6\,R_\star$), the opacity is controlled by electron scattering in the outer disk.

\begin{figure}[t!]
% \vspace*{-2.0 cm}
\begin{center}
 \includegraphics[width=3.5in]{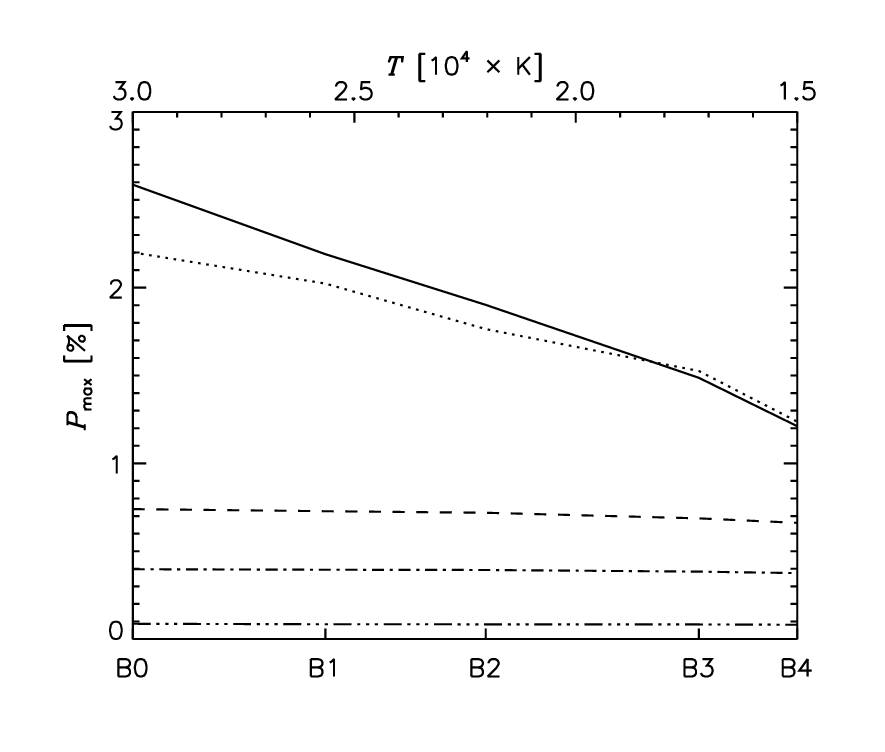} 
 \caption{$V$-band maximum level of polarization versus the effective temperature of the star for five different base densities and five spectral types. The inclination angle is 70$\degr$. From top to bottom: $8.4 \times 10^{-11} \rm g\,cm^{-3}$, $4.2 \times 10^{-11} \rm g\,cm^{-3}$, $8.4 \times 10^{-12} \rm g\,cm^{-3}$, $4.2 \times 10^{-12} \rm g\,cm^{-3}$ and $8.4 \times 10^{-13} \rm g\,cm^{-3}$.}  
   \label{polmaxxst}
\end{center}
\end{figure}

The resulting effects of the spectral type on the continuum polarization can be seen in Fig.~\ref{polxst}. The polarized spectrum goes for a relatively flat one for the B0 star (electron scattering dominated) to a steep one, with marked discontinuity at the \ion{H}{1} thresholds, for a B4 star at the density considered (bound-free dominated).
Another interesting aspect of Fig.~\ref{polxst} is the fact that the polarization level for a given wavelength increases with the effective temperature. 
This is seen more qualitatively in Fig.~\ref{polmaxxst} that shows how the $V$-band maximum polarization of a given model (that happens for an inclination around 70\degr, \citealt{wood96b}) varies with spectral type. For low densities, all models (including the ones for later spectral types) are electron scattering dominated, meaning that electron scattering opacity is larger than the bound-free opacity in the optical and near-IR continuum (see Appendix~\ref{appendix}, Fig.~\ref{loci_dens}, for an example of how the different opacities changes with changing disk density). This explains why the low density curve in Fig.~\ref{polmaxxst} is flat. 

As larger densities are considered, two effects concur for the strong dependence of the maximum polarization with spectral type. 
i) the relative contribution of the bound-free opacity to the total opacity increases, and eventually it will become the dominating source of opacity in the optical continuum (e.g., bottom panel of Fig.~\ref{tau_lambda}). Since absorptive opacities decrease the polarization level due to pre- and post-scattering absorption, it follows that the latter spectral types, that are more bound-free dominated, will have lower polarization levels. 
ii) the total scattering mass of the disk decreases for latter spectral types. For instance, in the high density B2 model of Fig.~\ref{polmaxxst}  the ionization fraction is about 99\%, while for the B4 model it is 91\%.  
These combined effects are so important that the maximum polarization level of the high density B0 model is about 2.5 times larger than the B4 model (Fig.~\ref{polmaxxst}).

In the above the relevance of continuum absorption to both the polarization level and the slope of the polarized continuum was discussed.
However, \citet{wood96b} showed that other processes, such as multiple scattering and occultation by the central star, also concur to define the shape of the polarized continuum. These latter processes are not discussed here, but they are all self-consistently included in our calculations \citep{car06}. In a recent paper \citet{2013ApJ...765...17H} also studied the polarization from Be star disks.
Based on an self-consistent treatment of the thermal structure of the disk, they performed a radiative transfer in a gaseous disk that feeds a Monte Carlo multiple scattering routine that provides polarization levels for different parameters. Even though a quantitative comparison was not attempted, their results shown on Fig.~1 are qualitatively similar to the ones we present in this paper.
%\rouge{Alex says:}

%\rouge{Really ??? or choose :  For a B2V star, the predicted polarization levels for different base densities \citep[Fig.~1,][]{2013ApJ...765...17H} are qualitatively similar to the ones we present in this paper. At the difference of the here-presented results, it is noteworthy they find qualitatively similar polarization levels for other spectral types.}

We conclude that the polarization spectrum carries valuable information about the physical conditions in the circumstellar disk. The polarization level at a given wavelength depends primarily on the total scattering mass of the envelope (number of free electrons), while the ''color'' of the spectrum (its slope or, equivalently, the size of the polarization change across the \ion{H}{1} ionization thresholds) depends on the \ion{H}{1} bound-free opacity, which is a strong function of both the disk density and the spectral type. These two polarization features (polarization level and ``color'') thus carry complementary information. This will be further explored in the next Section, particularly through the examples of the polarization in the $V$-band (polarization level), $P_{V}$, and the polarization level ratio across the Balmer discontinuity (polarization ``color''), $P_{\rm BD}$.

\section{Dynamical signatures in polarimetry}
\label{prediction}

In the last section, we set the stage for explaining the origin of polarized spectra from Be stars using a VDD model in steady state (constant mass decretion rate). However, Be stars are known to be highly variable, and a constant mass decretion rate is likely to be the exception rather than the rule. A much more common situation is a disk whose characteristics are time dependent, in response to variable disk injection rates.
To model the polarized signature of a time-dependent VDD, we follow the same approach used in Paper I. In that paper, the photometric variability was studied in detail and it was shown that predictions based on simple, yet realistic, mass injection rate scenarios agree quite well with the correlations observed for both shell and Be stars (we refer the reader to the discussion in \S~5 of Paper I). Here, our goal is to extend that study to the two polarimetric features described in \S~\ref{origin}: the polarization level in the $V$ band, and the color of the polarized spectrum as measured either by the polarization change across a \ion{H}{1} ionization threshold or by the slope of the polarized spectrum.

For the reader's benefit, we succinctly describe the procedures adopted here; full details are given in Paper I. Using the {\sc singlebe} code, different dynamical scenarios were built from a given coefficient of viscosity \citep[$\alpha$,][]{1973A&A....24..337S} and different histories of the mass injection into the disk: monotonic building-up and dissipation phases of the disk, periodic injection rates and episodic outbursts. From each hydrodynamic simulation, we obtained temporal series of surface density profiles for a given dynamical scenario. To transform these structural information into observables, we used the surface density profiles as input to the radiative transfer code {\sc hdust} for each epoch of the dynamical scenario we wanted to investigate. This 3-D Monte Carlo code produces a full spectral synthesis of a star+disk system with a non-LTE treatment of the level populations and gas temperature and produces output spectra for the Stokes parameters $Q$ and $U$. Unless mentioned otherwise,  the model parameters are the same as presented in Table 1 and Table 2 of Paper I. This  model that we call hereafter "the reference model" simulates a B2 star with a disk base density of  $8.4\times10^{-12} \rm g \,cm^{-3}$. Nevertheless, we also explored the base density and spectral type parameter space and the corresponding values are specified in the figure captions.

\subsection{Disk build-up and dissipation}
\label{builddiss}

\subsubsection{$V$-band Polarization and Balmer Discontinuity}

To understand the effects of different dynamical scenarios upon polarimetric observables we start with a case of an uninterrupted disk build-up and dissipation. The disk build-up is simulated using a constant mass injection rate. The simulation starts without any circumstellar material, and the disk is gradually fed with matter as time goes on. The dissipation is modeled starting with a fully-developed disk; mass injection is turned off, and the disk material slowly dissipates both inwardly (re-accretion onto the star) and outwardly. We refer the reader to Paper I for a detailed discussion on how the disk grows and dissipates, and the timescales involved.

\begin{figure*}[h!]
% \vspace*{-2.0 cm}
\begin{center}
 \includegraphics[width=3.5in]{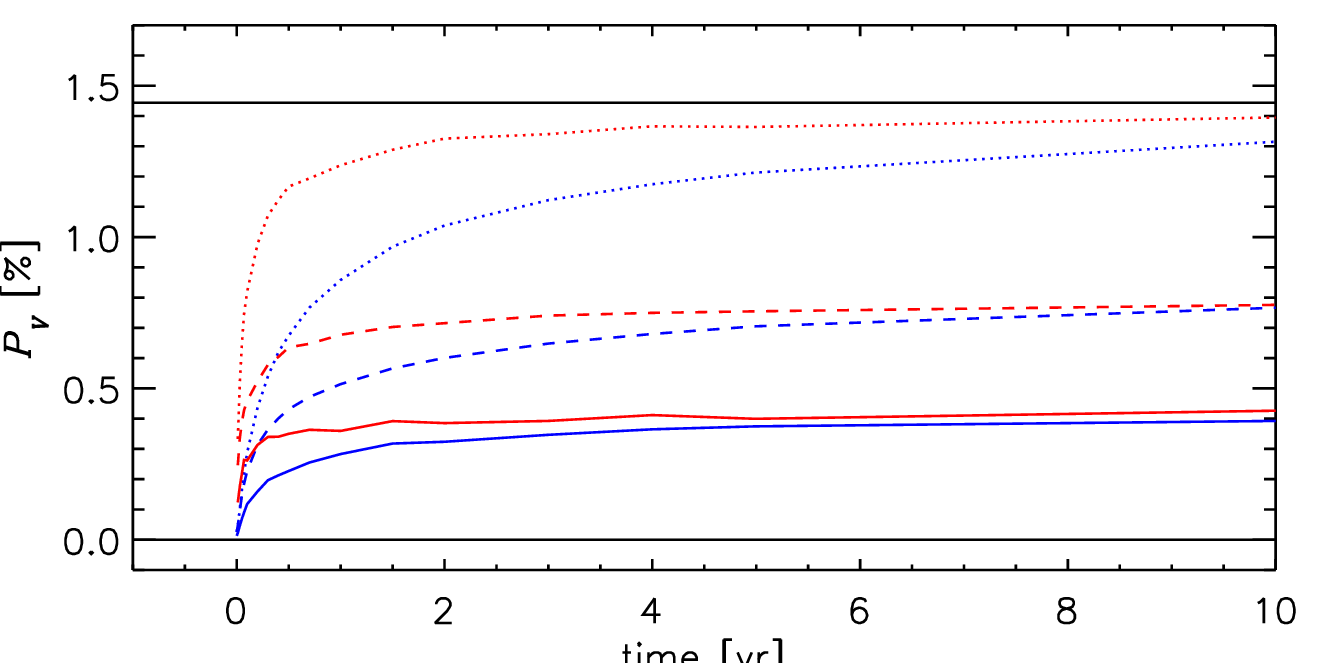}
  \includegraphics[width=3.5in]{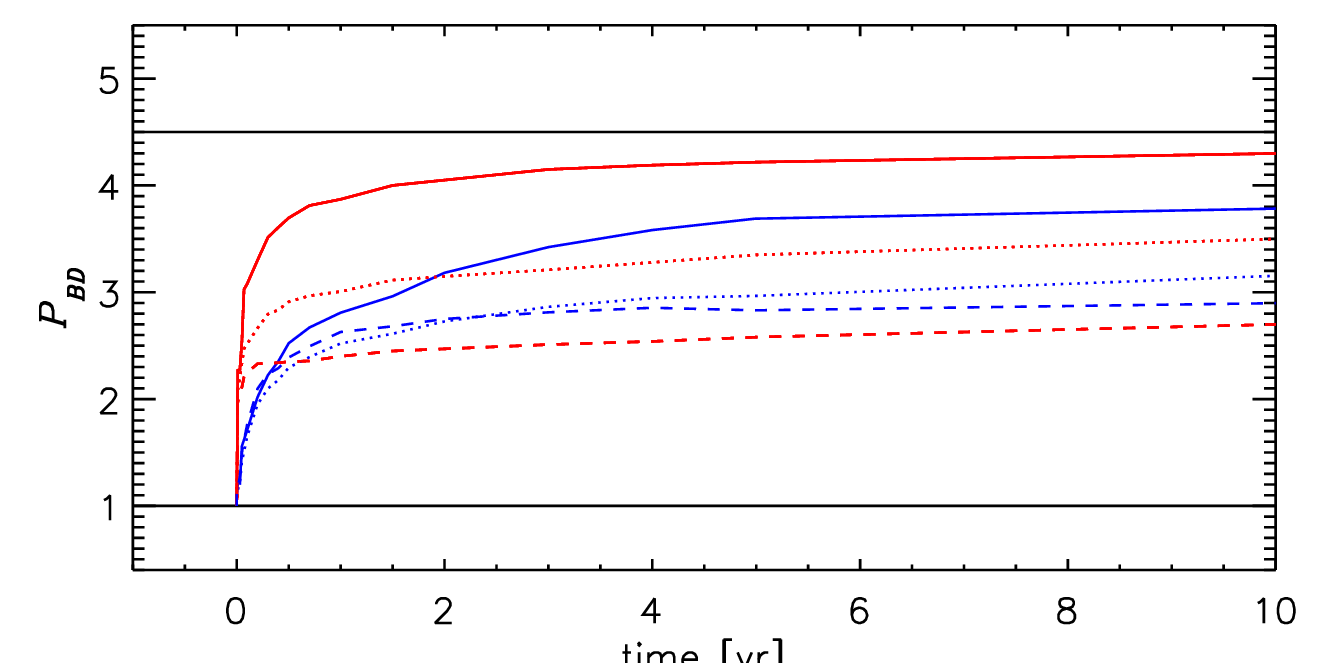}  
\includegraphics[width=3.5in]{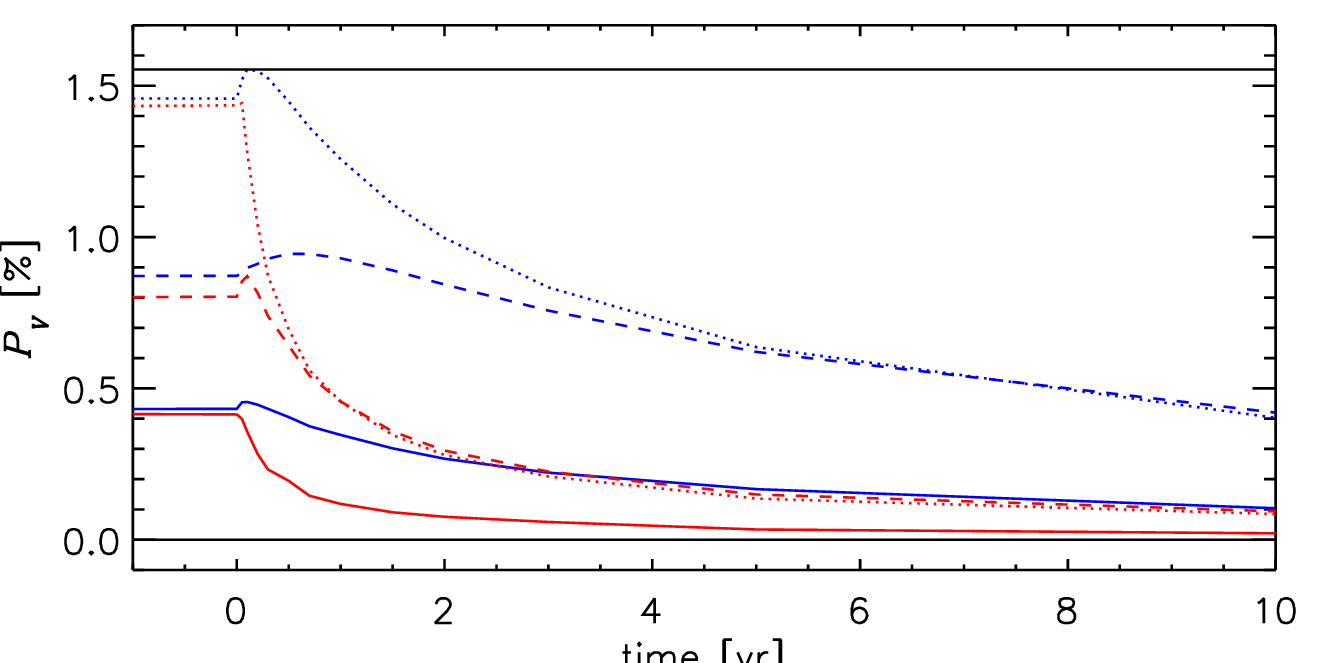} 
\includegraphics[width=3.5in]{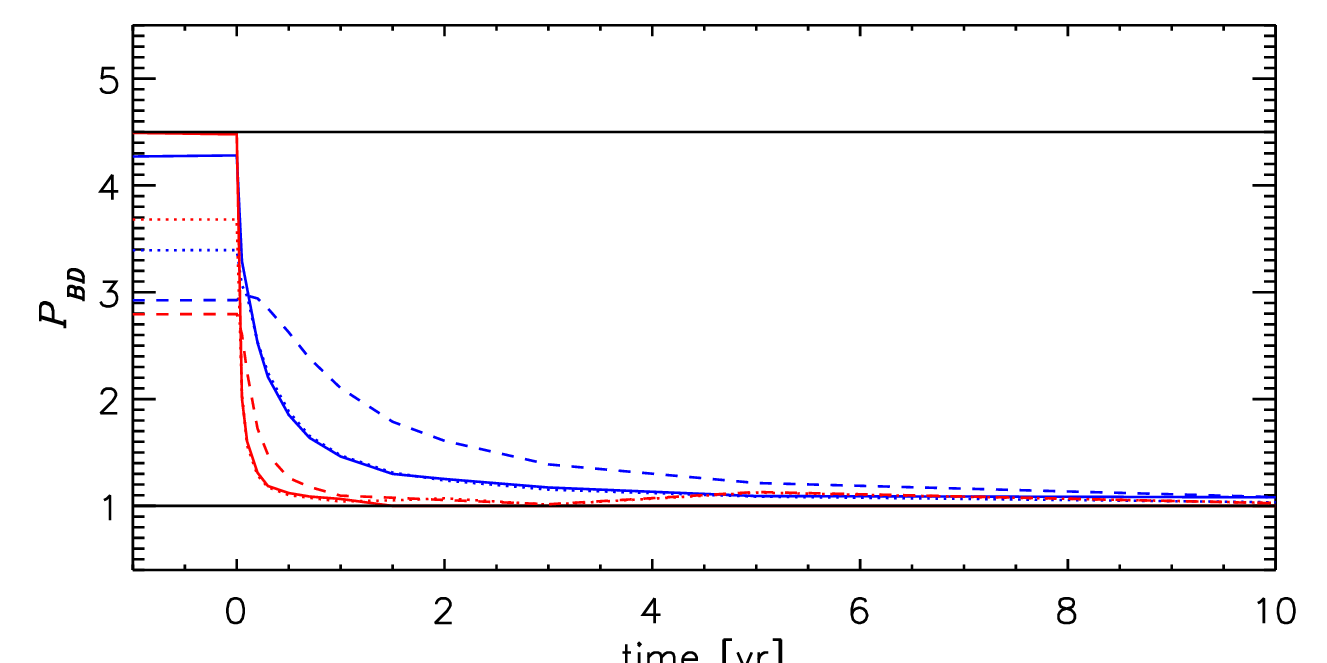} 
 \caption{Temporal evolution of $P_{V}$ (left) and $P_{\rm BD}$ (right) associated with disk build-up (top panels) and
dissipation (bottom panels). The solid, dotted and dashed lines represent the lightcurves for inclination angles of 30$\degr$(face-on), 70$\degr$ and 90$\degr$ (edge-on), respectively. The blue and red colors represent models for $\alpha$=0.1 and 1.0, respectively. The solid black lines indicate the maximum values for each scenario. These curves were obtained using the reference model parameters.}
   \label{growth_V}
\end{center}
\end{figure*}

To explore the two polarimetric features we mentioned at the beginning of the section, we plot in Fig.~\ref{growth_V} the temporal evolution of $P_{V}$ and $P_{\rm BD}$ defined as:
\begin{equation}
P_{\rm BD}= \frac{ P_{\lambda_{BD+}}}{ P_{\lambda_{BD-}}}
%D (\lambda)  =   P_{\lambda_{+}}  -  P_{\lambda_{-}}
\end{equation}
where $P_{\lambda_{BD+}}$ and $P_{\lambda_{BD-}}$ are the polarization levels estimated at higher and lower-wavelength intervals of the Balmer discontinuity, respectively. Results are shown for two $\alpha$ parameters and three inclination angles.

% V-band polarization stems from the total scattering mass of the disk (Figure \ref{loci}). 	

Understanding the build-up phase (upper panels of Fig.~\ref{growth_V}) is straightforward: the polarization level monotonically increases as the disk builds up, approaching a limit value after a time that depends on the $\alpha$ coefficient. This limit value is associated with the fact that even though viscous decretion disks never actually reach steady state, in the case of steadily building disks their surface density approaches a $r^{-2}$ profile when time goes to infinity (Paper I and references therein). The maximum of the polarization level does not happen for 90 $\degr$ but rather at around 70$\degr$ \citep[e.g.,][]{wood96b}. Indeed, at high inclinations, photons scattered in directions parallel to the disk plane are much more likely to be absorbed by the disk than photons scattered away from the disk. 

The dissipation (lower panels of Fig.~\ref{growth_V}) is characterized by an increase of the $P_{V}$ signal right after the mass injection has been set to zero (no more mass injected at the base of the disk). This somewhat counterintuitive behavior is explained by the fact that the (unpolarized) emission in the $V$-band decreases soon after the mass injection stops (see Paper I). However, the total polarized flux decreases less rapidly because it is produced in a bigger area in the disk \citep{rev2013}. This causes the polarization fraction to increase. When a sufficient fraction of the disk mass has been dissipated, there is less and less electrons available for Thomson-scattering, the polarization fraction naturally decreases as it is the case for $P_{\rm BD}$.
These variations are also naturally $\alpha$ and inclination angle dependent. Eventually, all the polarized signals reach the zero value, when the disk has dissipated all or almost all its scattering mass.  As shown in Fig.~\ref{growth_V},  $P_{\rm BD}$ grows and decreases faster than $P_{V}$ because the former quantity is a $\rho^2$-diagnostics (depending on the $\tau_{bf}$ opacity) whereas the latter varies linearly with the density (depending on the $\tau_{el}$ opacity.) Hence, $P_{\rm BD}$ responds much faster to changes in density.
To further illustrate this point, Tab.~\ref{tab:timescale} shows the timescales that $P_{V}$ and $P_{\rm BD}$ need to reach 95\% of their limit value for an uninterrupted 50 yearlong phase of building up (for $\alpha  = 1$).  The same timescales for the $V$-band magnitude are also listed for comparison (taken from Paper I).

  \begin{deluxetable}{cccc}
\tablecolumns{4}
\tablecaption{Time in years required to reach 95\% of the limit value\label{tab:timescale} for $\alpha  = 1$ }
\tablehead{\colhead{Inclination angle (deg)}  & \colhead{$T_{P_{\rm BD}}$} & \colhead{$T_{P_{V}}$} & \colhead{$T_{V mag}$} }
\startdata
30 & 1.4 & 1.9  &  2.8    \\
70 & 3.0 & 3.9  & $<\;$0.1 \\  
80 &  2.7 & 4.2 & 3.4   \\
85 &  1.0  & 4.8 & 7.1 \\
90 &  0.4 & 8.4 & 8.1 
\enddata
\end{deluxetable}

% And this decrease overcomes the increase previously explained that was due to a decreasing overall emission. \rouge{Is that clear enough Alex ?} 

%\paragraph{$V$-band magnitude vs P${_V}$ }
%
%From Figure 12 of Paper I and Figure \ref{loci} of the present paper, we can see that the excess magnitude and the polarization fraction in $V$-band stem from different locations in the disk. While the first quantity is produced in the first 2-3 $\,R_\star$, the second needs a larger pseudo-photosphere radius to fully develop (about $10\,R_\star$). These facts provide us with another interesting possibility of following viscous processes in Be disks such as dissipation, build-up or outburst phases because we have access to the local density formed in the inner part of the disk and the overall bulk density of the disk.
%
%Figure \ref{V_Pv} shows the $V$-band magnitude excess represented as a function of the $V$-band polarization for the same scenario presented in Figure \ref{period1}.
%
%
% \begin{figure}[h!]
%% \vspace*{-2.0 cm}
%\begin{center}
%% \includegraphics[width=3.5in]{mag_polV_period.eps} 
% \includegraphics[width=3.5in]{mag_polV_period2.eps} 
% \caption{$V$-band magnitude excess as a function of the $V$-band polarization for the same scenario presented in Figure \ref{period1}. Inclination angle is 30 (dots) and 70$\degr$ (dashes).}
%   \label{V_Pv}
%\end{center}
%\end{figure}

%Variation avec la densite -> le graph do Robbie (polar do)

\subsubsection{Polarization color diagrams (PCDs)}
\label{PCDintro}
From Fig.~12 of Paper I, we can see the excess magnitude in different bands stem from different locations in the disk. %While the first quantity is produced in the first 2-3 $\,R_\star$, the second needs a larger pseudo-photosphere radius to fully develop (about $10\,R_\star$).
As demonstrated in \S~\ref{origin}, the same apply for polarimetric features studied in this paper, which probes different disk regions. This offers the possibility of tracking the variations of the local density at different locations in the disk. This fact therefore provides us with an interesting possibility of following viscous processes in Be disks through different dynamical phases such as dissipation, build-up or outburst phases. 
%An interesting spectral region to observe with polarimetry is the Balmer Discontinuity (BD). Contrary to $P_{V}$, the polarization across the Balmer Discontinuity is a tracer of the innermost disk density.
Particularly, it is useful to look at  the correlation between the polarimetric features $P_{\rm BD}$ and $P_{V}$. This correlation was
 detected observationally in Be stars that underwent build-up and dissipation phases by \citet{2011ApJ...728L..40D}, which showed that the process of disk growth and dissipation is associated to a loop in a diagram that plots $P_{\rm BD}$ vs. $P_{V}$. We saw in Sect.~\ref{origin} that these quantities respond differently to density changes and are produced in different loci in the disk (although these loci do overlap). We can generalize the concept of this diagram by plotting other combinations of polarimetric observables that also represent a polarized spectrum ``color'' as a function of polarization level. In analogy to color magnitude diagrams (e.g. Fig.~22 of Paper I), this kind of diagram allows us to follow the evolution of Be star disks. We hereafter name this kind of diagram {\emph{polarization color diagrams} (PCD)}.

 \begin{figure}[h!]
%\centering
% \vspace*{-2.0 cm}
\begin{center}
 \includegraphics[width=3.5in]{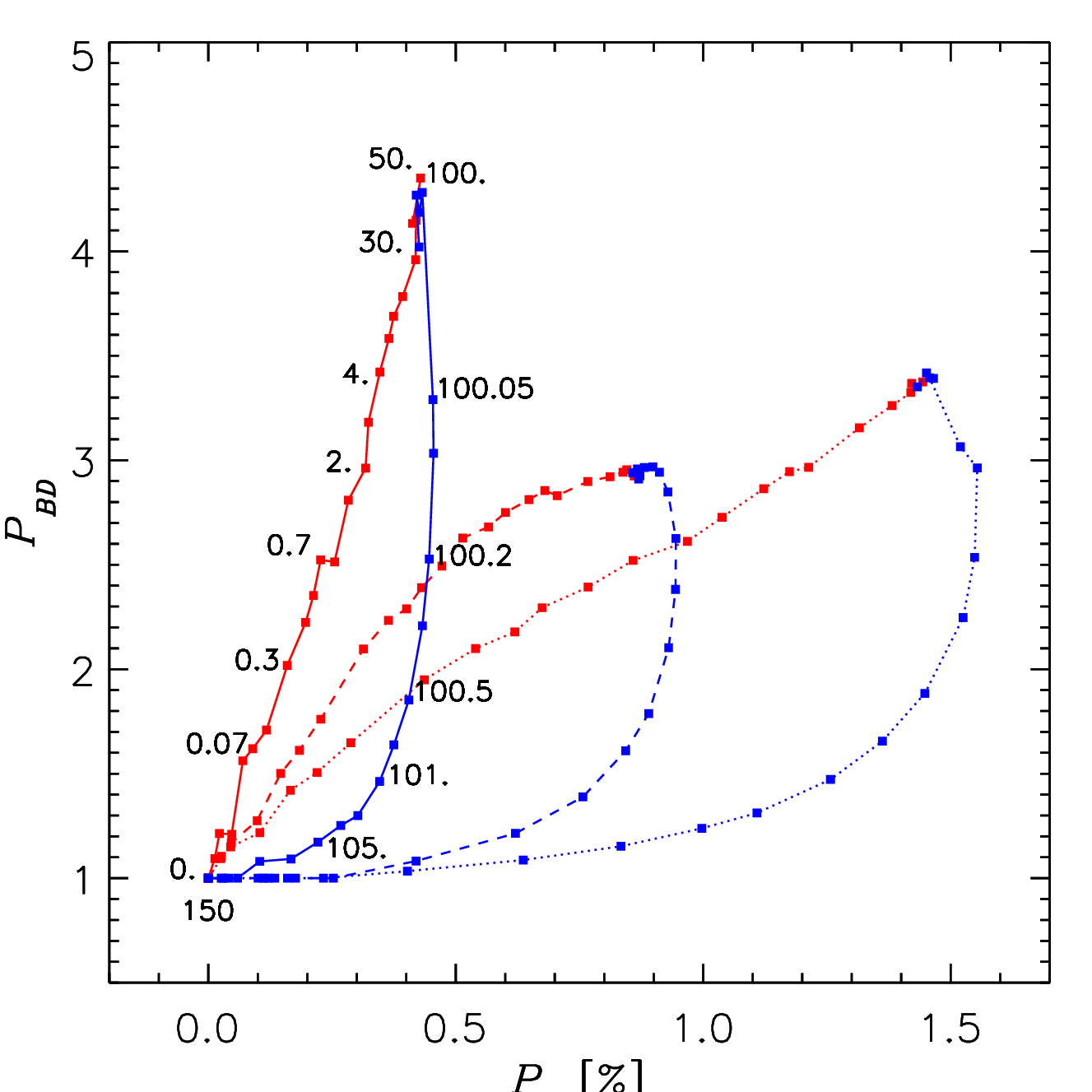}
 \caption{PCDs showing the process of disk growth and dissipation, involving a 100-year long building-up (in red) and 50 year dissipation (in blue). The results are shown for three inclination angles (solid line, 30$\degr$, dotted line, 70$\degr$ and dashed line, 90$\degr$). For reference, epochs are marked in years along the curve obtained for 30$\degr$ and $\alpha$=0.1. These curves were obtained using the reference model parameters.}
   \label{BDVD}
\end{center}
\end{figure}

Examples of PCDs are shown in Fig.~\ref{BDVD}. The formation of these loops can be described as follows. When the mass injection starts, $P_{\rm BD}$ increases from the value $1$ and faster than $P_{V}$ until it reaches a limit value as seen on Fig.~\ref{growth_V}. At the end of this first phase, the star has generated a large and dense CS disk. The formation path in the PCD corresponds to the upper part of the loops. When mass injection stops, the inner disk quickly reaccretes back onto the star; this causes a fast drop of $P_{\rm BD}$, which being a $\rho^{2}$-diagnostics, responds fast to the density variations. However, as described in the previous section, $P_{V}$ reacts more slowly to mass injections and increases a bit more until dissipation eventually makes its level drop. The curve then follows a track towards the bottom-left part of the diagram. What follows is a slow secular dissipation of the entire disk along which $P_{\rm BD}$ changes little and $P_{V}$ diminishes steadily. The detailed shape of the loop depends on the viewing angle as shown in Fig.~\ref{BDVD} and also on the value of $\alpha$.

\begin{figure}[h!]
% \vspace*{-2.0 cm}
\begin{center}
 \includegraphics[width=3.5in]{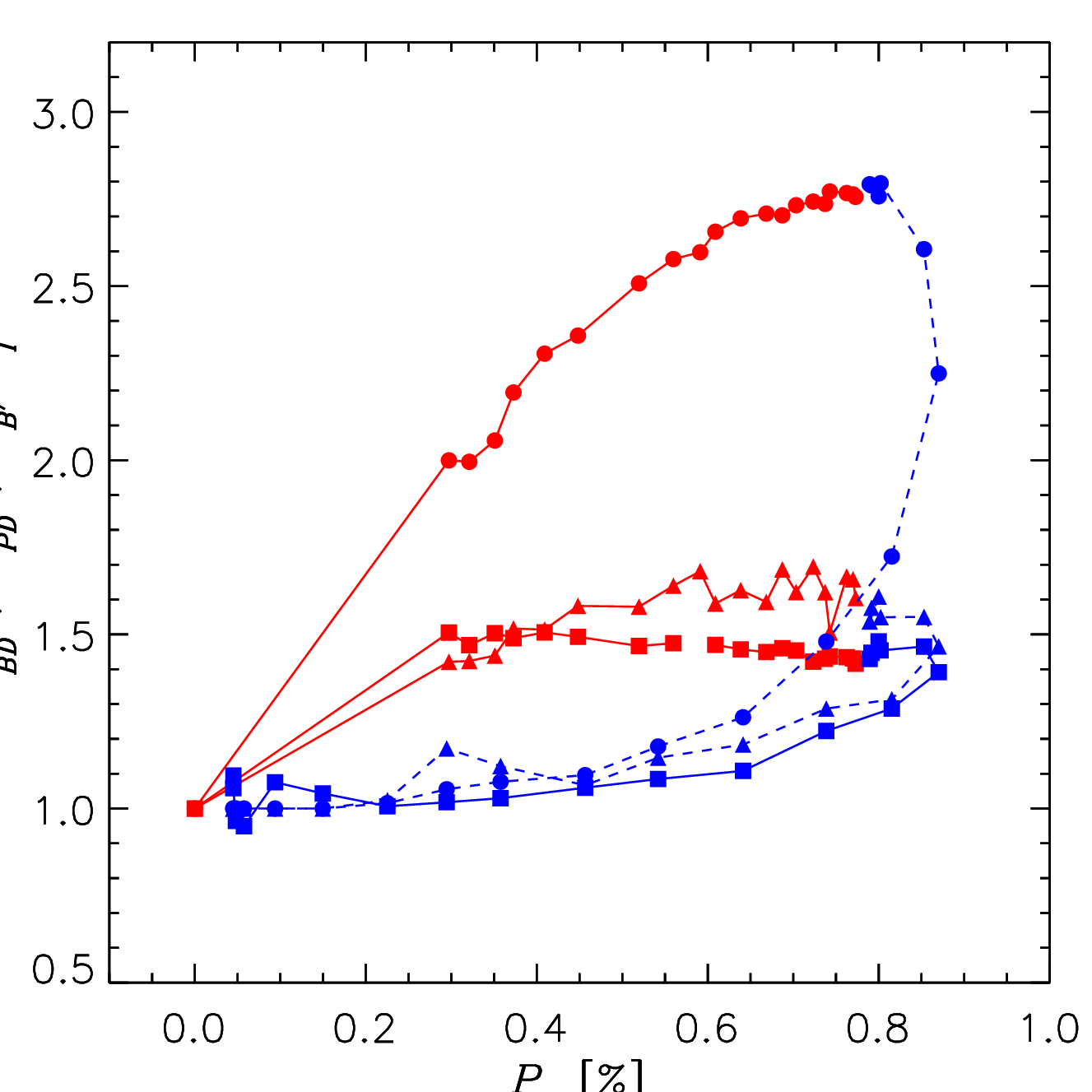}
 \caption{PCDs showing $P_{\rm BD}$ (filled circles), $P_{\rm PD}$ (squares) and $P_{B}/P_{I}$ (triangles) as a function of $P_{V}$. The red curve shows a 50 year building up and the blue curve a following 50 year dissipation phase. Inclination angle is 90$\degr$ and $\alpha$=1.0. These curves were obtained using the reference model parameters. }
   \label{col-pol-loop}
\end{center}
\end{figure}

To generate a PCD, we need to plot a quantity that measures the ``color'' of the polarized spectrum vs. the polarization level at some wavelength. Here, one can employ several quantities to measure the color; for instance, in addition to the polarization change in the Balmer discontinuity, other discontinuities can be used or, alternatively, the slope of the polarized spectrum can be measured directly from the ratio between the polarization values at different bandpasses.
An example of different diagrams for the same model can bee seen in Fig.~\ref{col-pol-loop} that compares the use of $P_{\rm BD}$ with the polarization change at the Paschen discontinuity ($P_{\rm PD}$) and the slope of the polarization spectrum as measured by the ratio between the polarization at the $B$ and $I$ bands (${P_{B}}/{P_{I}}$). All diagrams made according to this principle contain roughly the same information in the sense that they enable to track building-up and dissipation phases.
For simplicity, from now on by PCD we refer to the diagram $P_{\rm BD}$  vs. $P_{V}$.

\subsubsection{Diagnostic potential}

\begin{figure}[h!]
% \vspace*{-2.0 cm}
\begin{center}
 \includegraphics[width=3.5in]{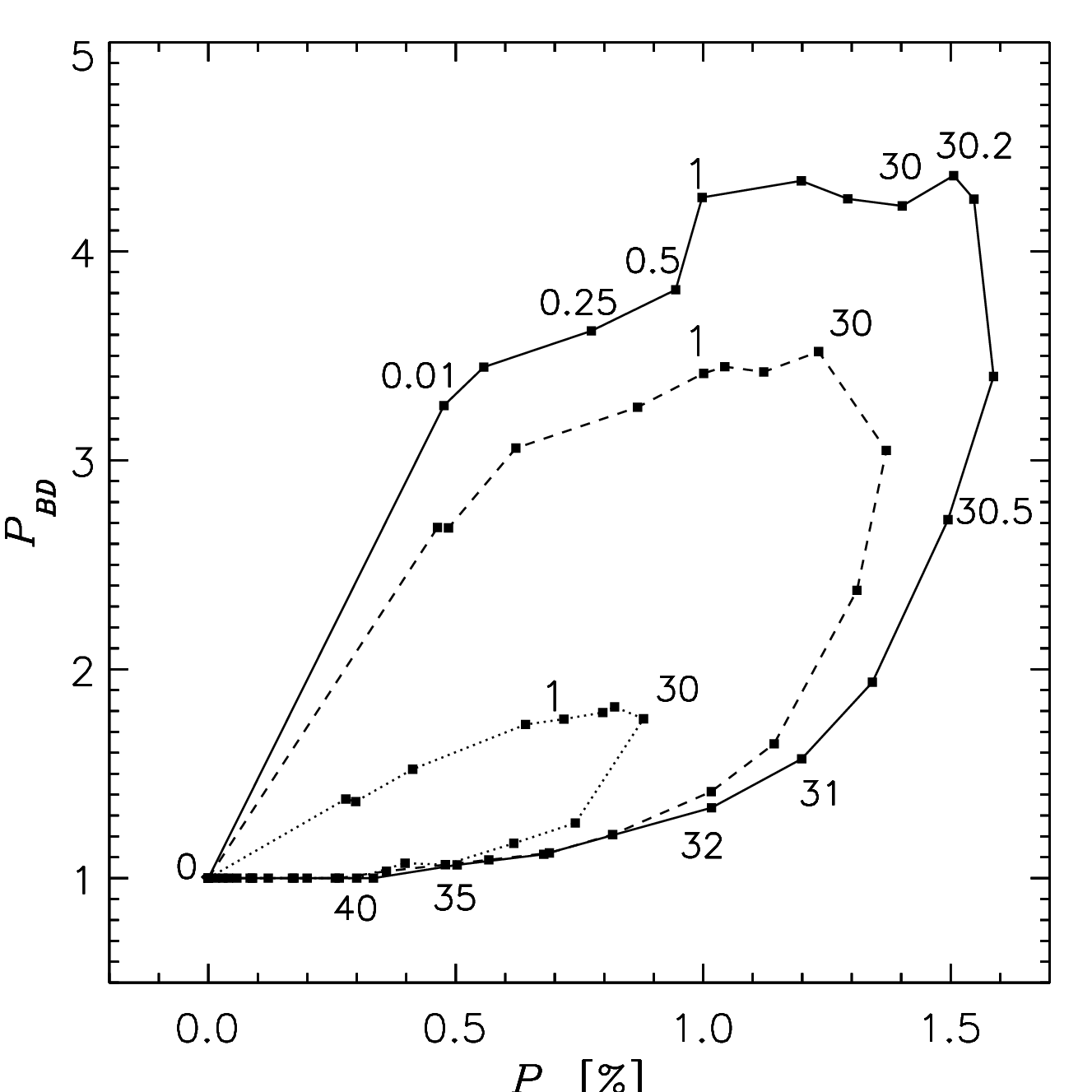}
\caption{PCD for the reference model with 3 different base densities:  $6 \times 10^{-11}\,\rm g\,cm^{-3}$ (solid line),  $3 \times 10^{-11}\,\rm g\,cm^{-3}$  (dashed line) and $1 \times 10^{-11}\,\rm g\,cm^{-3}$ (dotted line). The dynamical scenario involves a B3 star and 30 years of building-up followed by 30 years of dissipation. The inclination angle is 85$\degr$ and  $\alpha$=1.0. The epochs are marked and counted in years for the highest density model. The 1-year and 30-year epoch are marked for the three models.}
  \label{col-pol-loop2}
\end{center}
\end{figure}

Having described the two main features contained in the polarized spectrum (polarization level and color) and how these relate to the disk properties, we now explore the diagnostic potential of polarimetric observations of Be disks, with a focus on the usefulness of the PCDs. Figure~\ref{BDVD} already showed that the shape of the loop in the PCD is very sensitive to the inclination angle of the system. Moreover, PCDs are characteristic of the density at the base of the disk and can be used to infer this quantity via VDD modeling. Figure~\ref{col-pol-loop2} shows the different forms the PCDs can take depending on the density at the base of the disk.
The amplitude of the loop in the PCDs is a clear signature of the base density: the loop has a bigger extent for a higher density, whatever the inclination angle. If the disk undergoes an almost constant building-up phase, the top region of the PCDs is reached in less than a year (for a viscous coefficient $\alpha = 1$). Provided a few measurements typically spaced by one or two months and with $\sim$0.1\% uncertainties, Fig.~\ref{col-pol-loop2} shows that it is quite straightforward to disentangle between base densities that are different by a factor of 3 (except may be at the end of a long-term dissipation phase). We also note that the slope of the PCDs changes with density. For low densities the slope is smaller and the loop less broad, while for large densities the slope increases and the loop gets broader. This is a result of the different weight that \ion{H}{1} opacity plays in each model.

\begin{figure}[h!]
% \vspace*{-2.0 cm}
\begin{center}
 \includegraphics[width=3.5in]{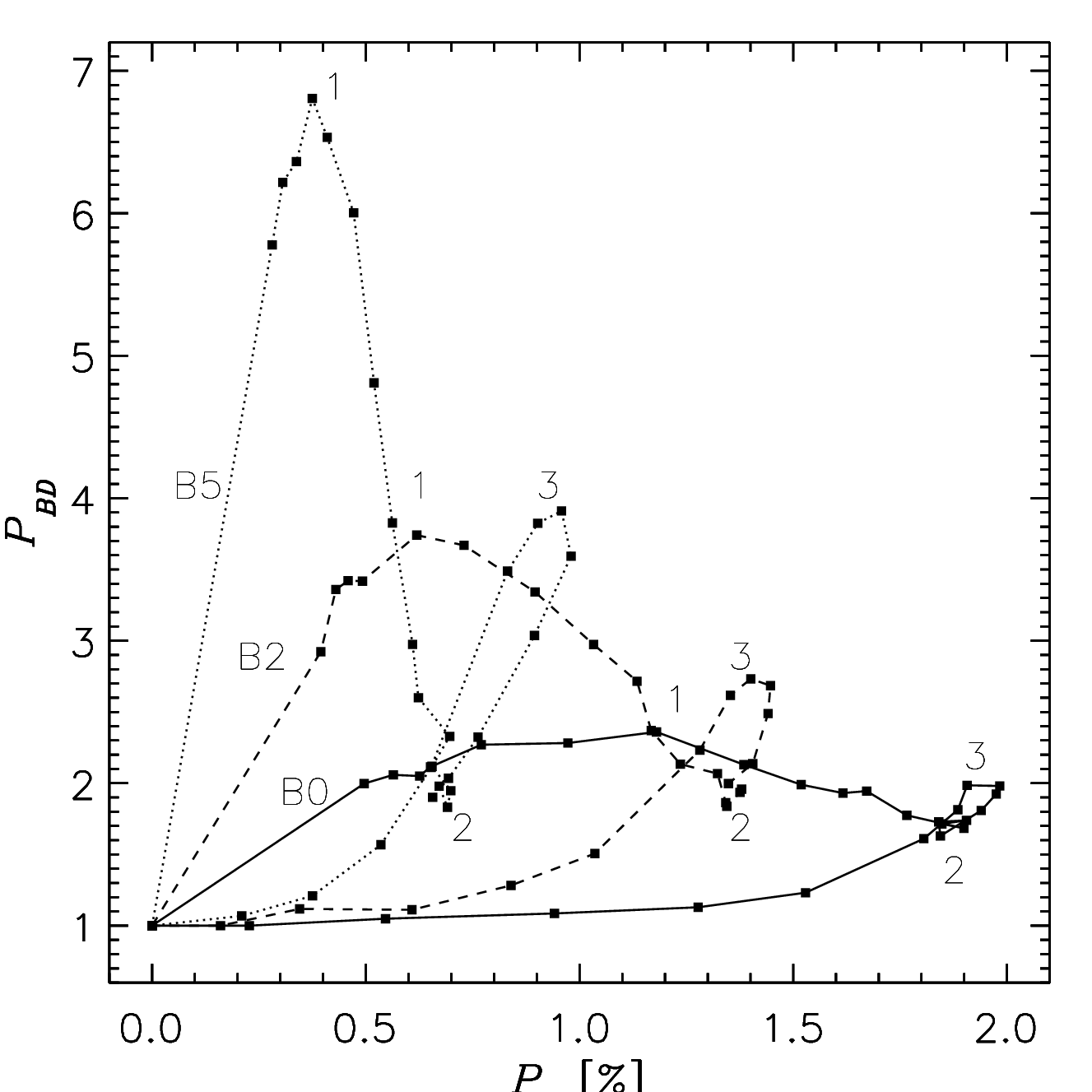}
 \caption{PCDs for 3 different spectral types (solid for B0, dashed for B2 and dotted for B5) with a dynamical scenario compound of 30 years of building-up followed by 30 years of dissipation. The base density is  $1\times10^{-10}$$\,\rm g\,cm^{-3}$. The inclination angle is 90$\degr$ and  $\alpha$=1.0. Numbers indicate three reference times that help describing the disk evolution in the text. Whereas $1$ and $3$ represent different epochs depending on the spectral type, the number $2$ always marks the stop of mass injection in the disk at $t=30$ years. }
  \label{col-pol-loop4}
\end{center}
\end{figure}

Interestingly, PCDs exhibit a secondary loop for very high density models ($> 1 \times 10^{-10}\,\rm g\,cm^{-3}$) seen close to equator-on (Fig~\ref{col-pol-loop4}).
For all the PCDs presented so far, the tip of the loop roughly corresponded to the end of the disk build-up phase (for instance, see Fig.~\ref{BDVD}). This is easily understood from the fact that the higher the density, the higher the polarization change across the Balmer jump (e.g., Fig.~\ref{tau_lambda}). However, for the models shown in Fig.~\ref{col-pol-loop4}, the tip of the loop (labeled point 1) no longer corresponds to the end of the build-up phase (labeled point 2). 
This phenomenon is characteristic of high inclination angles ($i>80\deg$) and the most likely explanation is a rather complex interplay between pre-and post-scattering absorption. 
Indeed, the two effects cause both a decrease in the polarization value and a change in the color of the polarized spectrum. However, pre-scattering absorption depends on the conditions of the innermost part of the disk, whereas post-scattering absorption follows the conditions at much larger radii. 
We thus explain Fig.~\ref{col-pol-loop4} as follows:
For the onset of the disk formation to point 1, the loops follow a normal behavior: as time passes $P_{\rm BD}$ grows as a result of the density increase everywhere in the disk. As the disk continues to grow and to get denser, post-scattering absorption starts to play a role in lowering $P_{\rm BD}$ (point 1 to 2). The track between point 1 and 2 in Fig.~\ref{col-pol-loop4} thus indicates the growing role of the outer disk in absorbing the polarized flux coming from the innermost parts of the disk. When mass injection stops (point 2), $P_V$ increases for reasons already explained in \S~\ref{builddiss}, and $P_{\rm BD}$ also increases due to the smaller post-scattering absorption. 
 At point 3, post-scattering absorption becomes negligible and, at the same time, the $P_{\rm BD}$ also drops due to the lack of density. This is why the $P_{\rm BD}$ level at the top of the secondary loop (point 3) is always smaller than the level at the top of the primary loop (point 1). Finally, the PCD follows a normal track towards $P_{\rm BD}=1$ and $P_V=0$, characteristic of an emptying disk. Moreover, since this post-scattering absorption changes the color of the polarized spectrum, the phenomenon we just described is more important for late spectral type stars where the polarization color is more pronounced. From the observational point of view this example shows that it is not possible to directly associate a value of $P_{\rm BD}$ to a disk density scale, at least for close to edge-on viewing.

\begin{figure}[h!]
% \vspace*{-2.0 cm}
\begin{center}
 \includegraphics[width=3.5in]{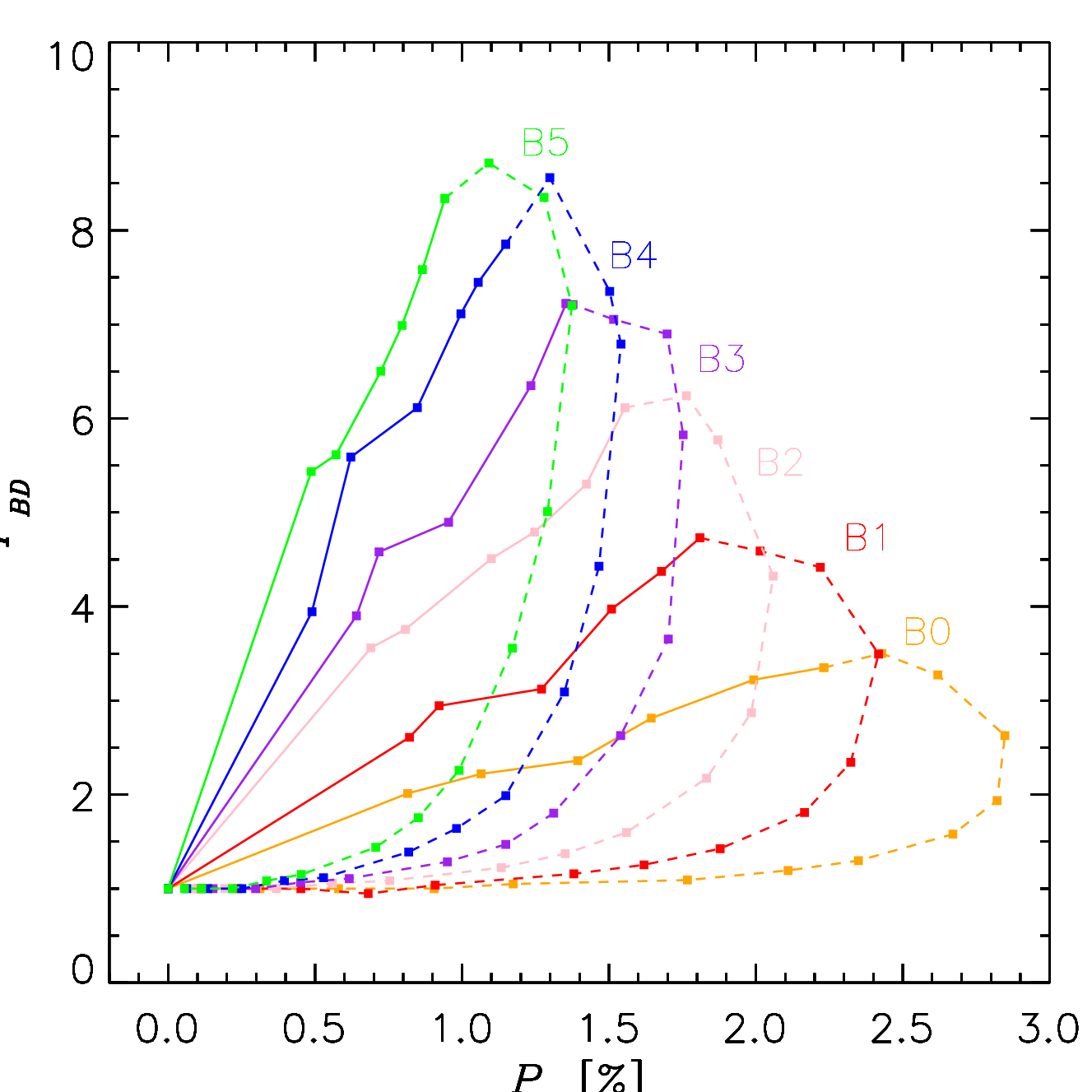}
\caption{PCD for a central star of different spectral types, as indicated. The scenario is the same as for Fig.~\ref{col-pol-loop2}. The solid lines mark the build-up phase whereas the dashed lines represent the dissipation phase. The base density is  $6\times10^{-11}\,\rm g\,cm^{-3}$. The inclination angle is 70$\degr$ and $\alpha$=1.0.}
  \label{col-pol-loop3}
\end{center}
\end{figure}

Fig.~\ref{polmaxxst} showed how the maximum polarization of a given model depends on $T_{\rm eff}$. A strong dependence of the PCD on the central star spectral type is thus to be expected. On Fig.~\ref{col-pol-loop3}, we can see that the spectral type impacts the PCD in various ways. Firstly, the range of  $P_{V}$ values plotted in the PCD increases with increasing $T_{\rm eff}$ (in agreement with Fig.~\ref{polmaxxst}). Secondly, the slope of the PCD increases with decreasing $T_{\rm eff}$. This again can be understood in terms of the relative importance of electron scattering opacity vs. \ion{H}{1} opacity.

%The shape of the PCDs is also characteristic of the spectral type of the central star. Figure~\ref{col-pol-loop3} represents how this parameter impacts a PCD.

As seen above, the higher the effective temperature, the lower the bound-free opacity for a given density. Also, the lower the bound-free opacity, the smaller the polarization change across the Balmer discontinuity. Therefore, the slope of the loop in the PCD is a good tracer of the bound-free opacity.
In view of the results shown in Figs.~\ref{col-pol-loop2} and \ref{col-pol-loop4}, the diagnostic potential of the PCD as a measure of the disk density depends critically on a good determination of the spectral type of the central star.
 
% Those facts therefore establish the theoretical possibility of estimating the spectral type of the central star using PCDs. 

\citet{2013ApJ...765...17H} studied the shape of the PCD loops using add hoc models that simulated the disk growth and dissipation by simply changing the inner and outer radius of the disk while keeping the density slope fixed to its steady state value ($n=3.5$). A quantitative agreement between this work and their results is not to be expected due to the different disk models used but also because their definition for $P_{\rm BD}$ is based on the difference between polarization levels before and after the Balmer discontinuity instead of the ratio (definition used in the present work).

%However, while their models do produce loops in the PCD, in agreement with \citet{2011ApJ...728L..40D} and this work, some important differences are seen. For instance, \citeauthor{2013ApJ...765...17H} find that the size of the polarization change across the Balmer discontinuity (or, equivalently, the slope of the loop) increase for earlier spectral types, while in this work the opposite behavior is found (Fig.~\ref{col-pol-loop3}).

%\rouge{ CHECKING: the opacity is lower for earlier ST at 0.364 microns (from the analytical part, figure 2), why do decretion models show a lower $P_{\rm BD}$ then ?}

\subsection{Periodic mass loss rate}

The previous section dealt with the case of a disk forming when none exists and the dissipation of a fully formed disk. This situation has been already observed in several Be stars. Some examples include $\pi$\,Aqr \citep[a well-documented disk dissipation that started in 1986 can be found in][]{2010ApJ...709.1306W} and the disk growth of $\omega$\,Ori in the eighties \citep{1988ApJ...325..784S}. In between these limiting cases, most Be stars display either an irregular variability, or, what is more rare, a quasi-cyclic variation of the light curve \citep{2008A&A...478..659S}. The best-studied example of the latter case is $\omega$\,CMa, that displays 2--3\,yr long outbursts separated by 4--5\,yr of quiescence \citep{2003A&A...402..253S}. In this section we explore what is the polarization signature expected when the mass injection is turned on and off periodically.
 \begin{figure}[h!]
% \vspace*{-2.0 cm}
\begin{center}
 \includegraphics[width=3.6in]{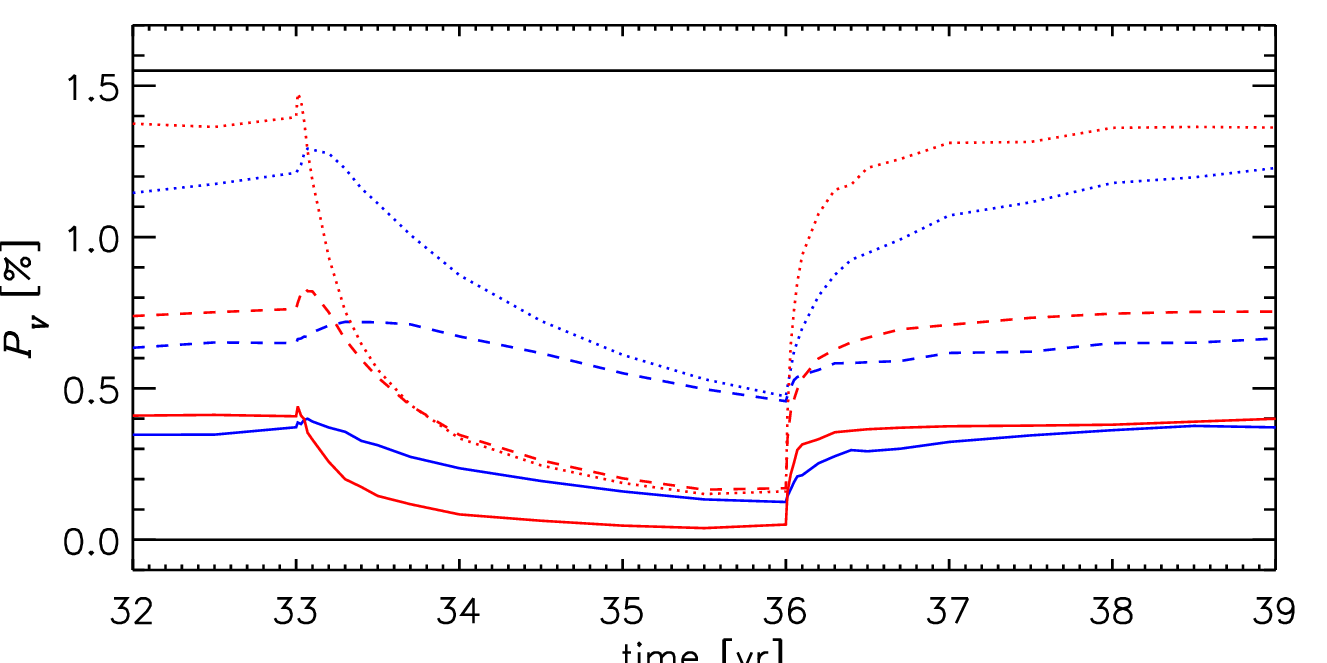}
 \includegraphics[width=3.6in]{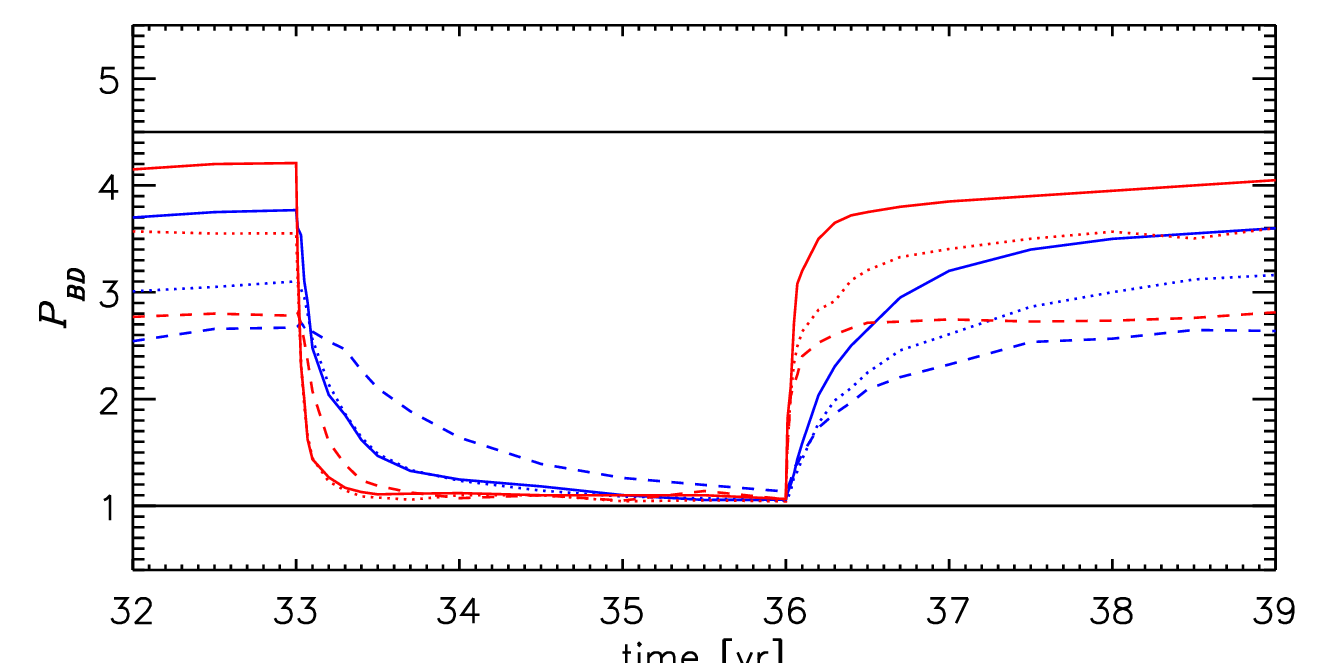}
 \caption{Same as for Fig.~\ref{growth_V} for the 6th cycle of a periodic mass injection scenario with period of 6 years and a duty cycle of 50\%. The solid black lines indicate the limit values that are reached for uninterrupted build-up and dissipation phase.}
   \label{period1}
\end{center}
\end{figure}

\begin{figure}[h!]
%\centering
% \vspace*{-2.0 cm}
\begin{center}
 \includegraphics[width=3.5in]{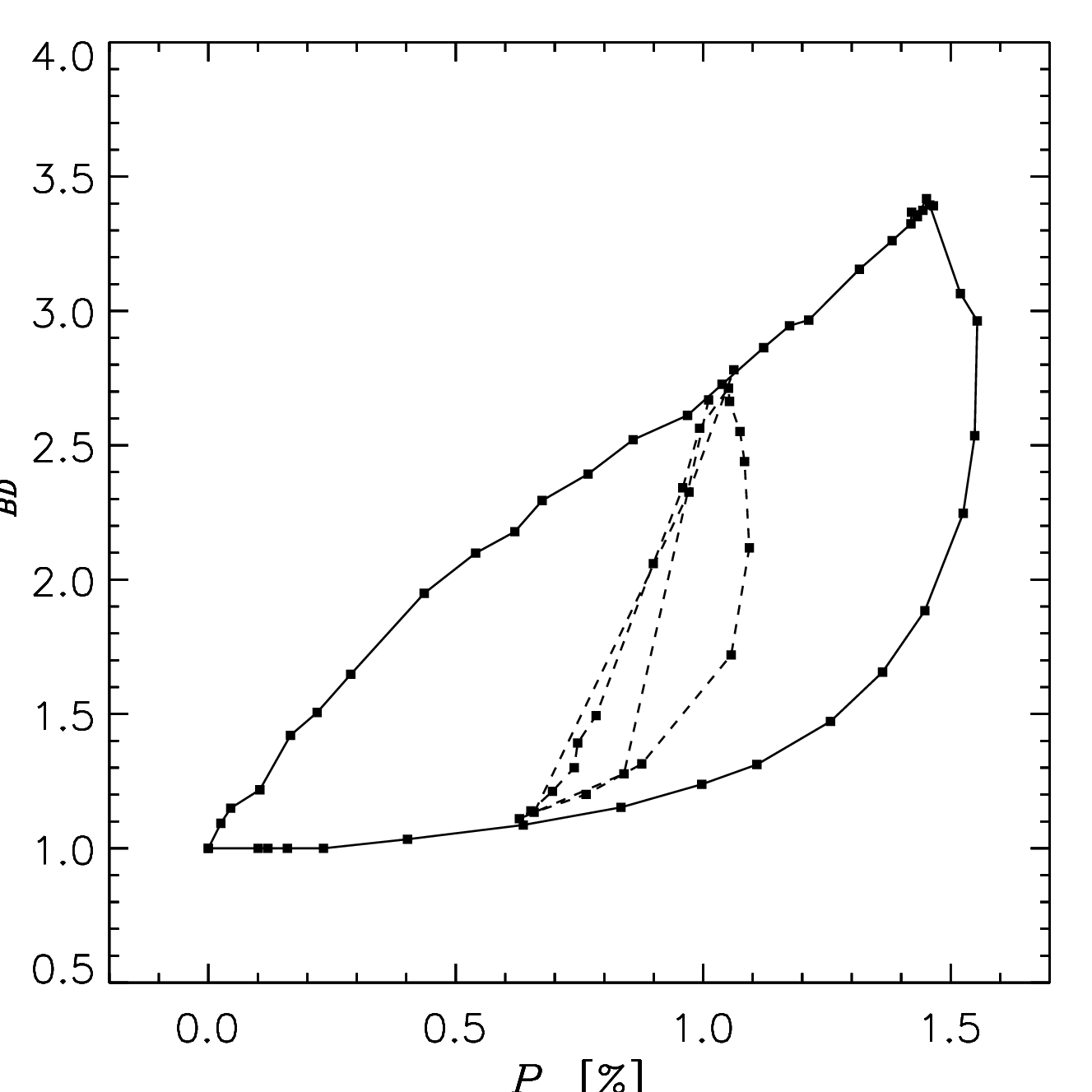}
 \caption{PCD at 70$\degr$ of inclination angle  and with $\alpha = 0.1$ for the same dynamical scenario shown in Fig.~\ref{BDVD} (in solid line) and for another one where the mass injection rate is periodically turned on and off every one year (in dashed line).}
   \label{BDVD2}
\end{center}
\end{figure}

A periodical scenario is defined by three parameters: the mass injection rate, the period, and the duty cycle, which indicates the fraction of time in each cycle where mass injection rate is larger than 0 (see Paper I for details). Figure~\ref{period1} shows the temporal evolution of  $P_{V}$ and $P_{\rm BD}$ for a periodic mass injection scenario (6--yr period and 50\% duty cycle, meaning that mass is continuously injected into the disk for three years every six years). The main difference between this case and the scenarios seen in \S~\ref{builddiss} is that the disk doesn't totally build and dissipate. The lowest values of $P_{V}$ and $P_{\rm BD}$ at the end of quiescence are therefore different from $0\%$ and $1$, respectively, and the maximum values are less than for an uninterrupted building-up phase (Fig.~\ref{growth_V}).
Figure~\ref{BDVD2} compares the tracks in the PCD of a periodical scenario (1-yr period, 50\% duty cycle, dashed curve) to that of a fully formed disk (formation from no previous disk followed by full dissipation, solid curve). The latter case forms a \emph{closed loop with maximal extent} because at the end of the build-up phase the disk density approaches a limit value (Paper I), and at the end of the dissipation the disk matter has been almost completely lost.
In the case of a periodical mass injection, however, there are three main differences: i) the cyclic mass injection prevents the disk density to reach its limit value,  ii) at the end of quiescence the disk matter has not yet been fully lost, and iii) a given cycle starts with matter already accumulated in the previous one, so the total disk mass of each successive cycle, at a given phase, is always larger than the previous ones. Observationally, we already saw on Fig.~\ref{period1} that these differences imply that the variations of $P_{V}$ and $P_{\rm BD}$ will be of a smaller amplitude but it also results in a variation of these curves from cycle to cycle. This last points depends on the parameters of the scenario and is mainly visible for low-period scenarios (i.e. about a year or less for $\alpha = 1.0$). This is well illustrated in Fig.~\ref{BDVD2}: the loops in the PCD for a periodic scenario never close, and they are of a smaller amplitude than the loops for fully formed and dissipated disks in which they are confined. %This demonstrates the diagnosis potential of this type of plot to infer periodic scenarios in Be stars.
 
%With a limited time coverage of observed measurements, it is very likely that the here-presented synthetic polarization curves will not be constrained enough and a same dataset could equally be modeled by different dynamical scenarios and star/disk parameters. The solution remains in the observation at more wavelengths to break this degeneracy.

\subsection{Episodic mass loss rate}
 \label{episodic}
The most common type of variability observed in Be stars is of an irregular nature, which means that the mass injection rate varies with time in a complicated way. One extreme example are the so-called ``flickering activity'', which is related to short-term variations seen in several observables. Examples of flickering activity can be found in \citet{1998A&A...333..125R} ($\mu$\,Cen, photometry and spectroscopy) and \citet{2007ApJ...671L..49C} ($\alpha$\,Eri, polarimetry). The ``flickering'' is attributed to an enhanced mass injection rate (outburst) that lasts from a few days to several weeks.

  \begin{figure}[h!]
%\centering
% \vspace*{-2.0 cm}
\begin{center}
 \includegraphics[width=3.5in]{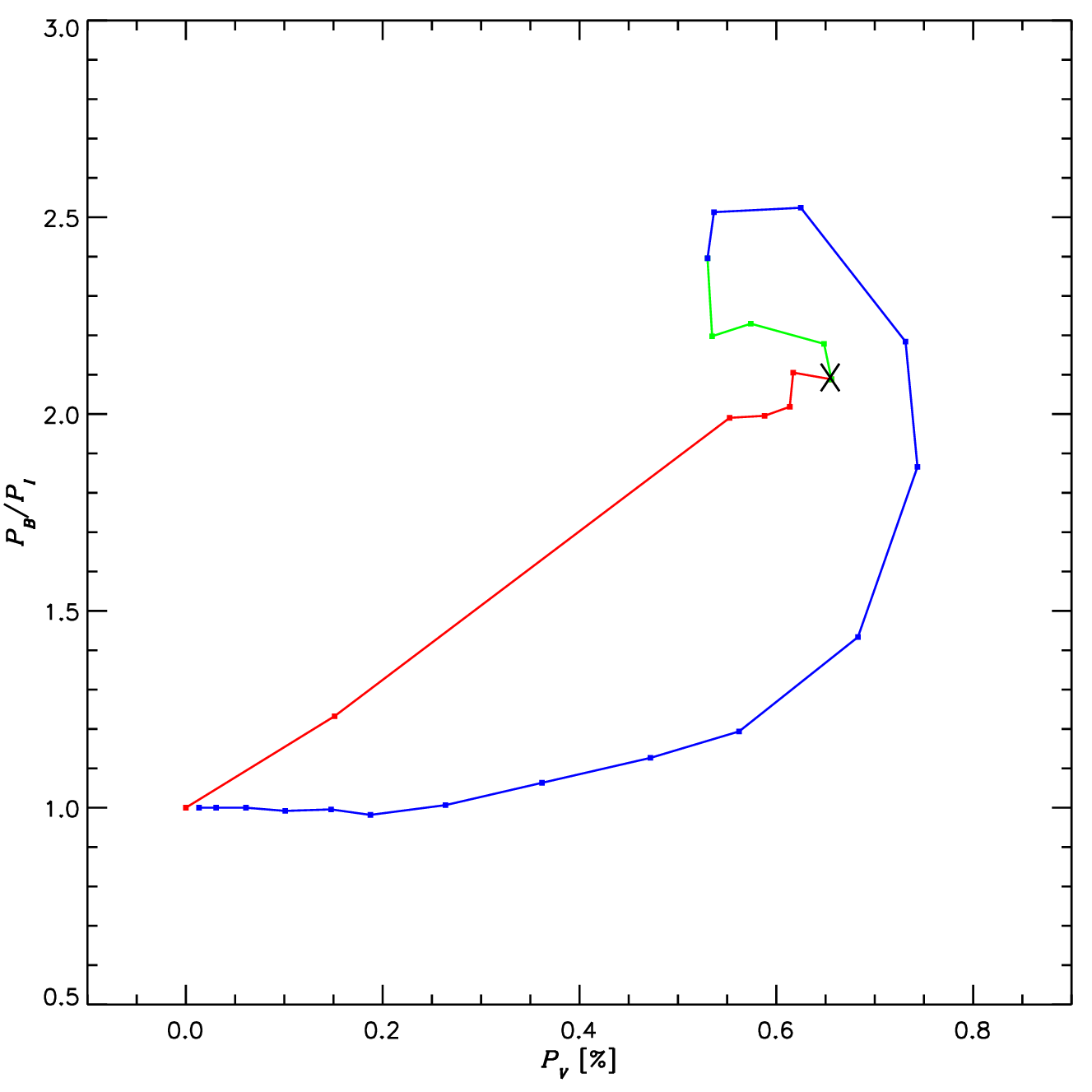}
 \caption{PCD showing $P_{B}/P_{I}$ as a function of $P_{V}$ for a 0.2 year outburst ($\dot{M}= 3.3\times10^{-8} M_{\odot}$/year, green curve) following a 20 year-long ($\dot{M}=1.6\times10^{-8} M_{\odot}$/year, red curve) stable period of building-up. The outburst is then followed by a 30 year dissipation (blue curve).  The X point marks the start of the outburst. The parameters are those of the reference model (B2 star). The inclination angle is 39 \degr and $\alpha$=1.0.  %The solid and dashed lines represent inclination angles of 70 and 90$\degr$, respectively.
 }
   \label{BDVD3}
\end{center}
\end{figure}
 
To illustrate what are the effects of an outburst in the polarimetric features, Fig.~\ref{BDVD3} shows a PCD for a dynamical scenario involving a 20-year long building-up phase followed by a 0.2 year outburst with a mass injection rate twice higher than for the previous phase. The outburst is then followed by a 30 year dissipation phase. The first 20 years of disk formation displays the same signature in the PCD as seen above (red line in Fig.~\ref{BDVD3}).
After the mass injection rate doubles at $t=20$\,yr, the disk density gradually increases inside-out (see Fig.~11 of Paper I for a description on how the disk density evolves with time in this scenario). The signature of the outburst (green curve in Fig.~\ref{BDVD3}) in the PCD is quite curious: while $P_{B}/P_{I}$ stays essentially the same,  $P_{V}$  initially \emph{ decreases} between epochs 20.01 and 20.1 years. This can be understood in terms of the sudden increase in the disk continuum emission (Fig.~21 of Paper I) as a result of the enhanced densities in the inner disk. After the disk total mass adjusts to the new mass injection rate ( $\dot{M}=1.6\times10^{-8} M_{\odot}$/year at epoch 20 years or so), then $P_{V}$ increases again. After the end of the outburst, mass injection is turned off ($t=20.2$\,yr) and the model follows a dissipative path in the PCD (blue curve), as seen before. The outburst thus adds an extra outgrowth to the loop compared to a dynamical scenario without any outburst and should be quite well-identified provided a sufficient time-coverage of polarimetric observations.

\section{Comparison of the predictions with observed data}
\label{comp}
   
%   WIS2010 detailed how the gradual disk-loss episodes of 60Cyg and ¹ Aqr proceeded over a timescale of ?1000 and ?2400days, respectively, and how these events were temporarily stalled by several polarimetric outbursts (Figure 1, left panels; see also WIS2010)

Long-term polarimetric observations of the Be stars 60 Cygni and $\pi$ Aquarii were obtained with the spectropolarimeter HPOL and revealed several year long disk-loss episodes interrupted by outbursts \citep{2010ApJ...709.1306W}. PCDs made out of these measurement showed loop patterns whose general shape was generally well described with viscous disk build-up and dissipation phases \citep{2011ApJ...728L..40D}. More recently, \cite{Draper2013} presented a polarimetric follow-up of 9 Be stars among which some of them exhibited very characteristic PCD patterns, similar to the ones we showed in the last section.

In this section, we compare our theoretical predictions with PCDs obtained from HPOL measurements for 60 Cygni, $\pi$ Aquarii, and $\psi$ Persei. We refer the reader to \cite{Draper2013} for further details regarding the observational data.
Moreover, we also discuss polarimetric measurements of $\delta$ Scorpii made with the IAG polarimeter at the Pico dos Dias Observatory \citep{bed2012}.

%  \begin{figure}[h!]
%\begin{center}
% \includegraphics[width=2.7in]{60cygPiaqr.eps} 
% \caption{Observed PCDs of 60 Cyg and $\pi$ Aqr from \cite{Draper2013}.}
%   \label{cygaqr}
%\end{center}
%\end{figure}

\begin{figure*}
\begin{center}
 \includegraphics[clip,width=2.2in]{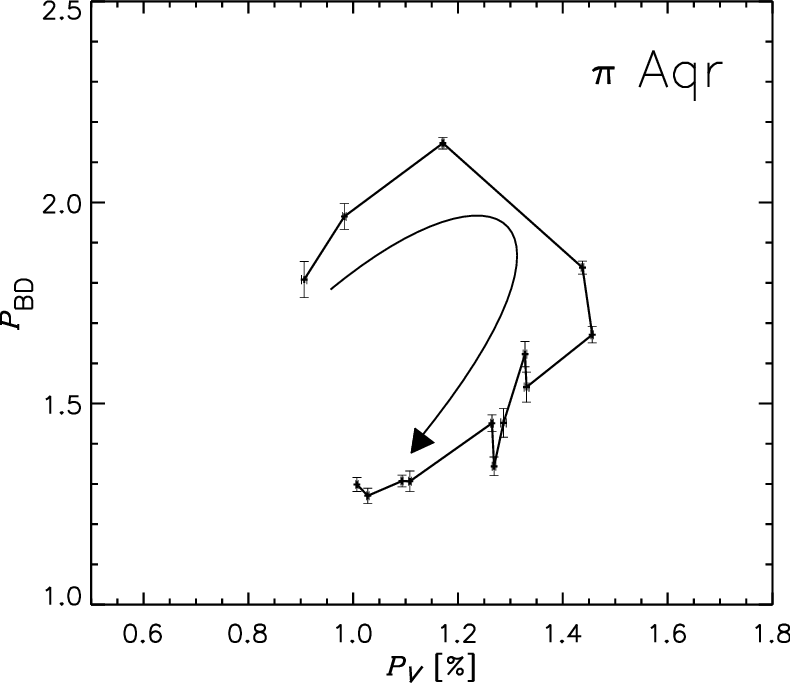}%\label{piaqr}
 \includegraphics[clip,width=2.2in]{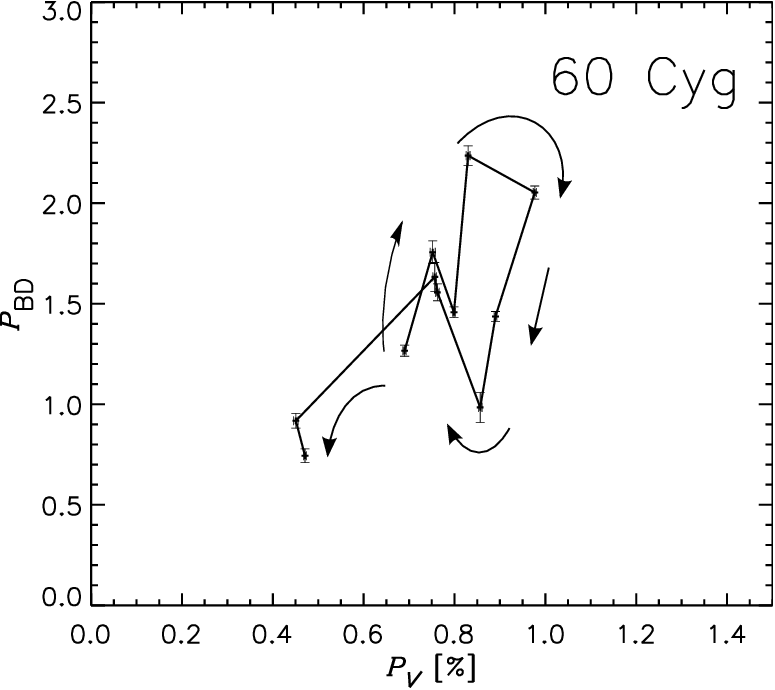}%\label{60cyg}
 \includegraphics[clip,width=2.2in]{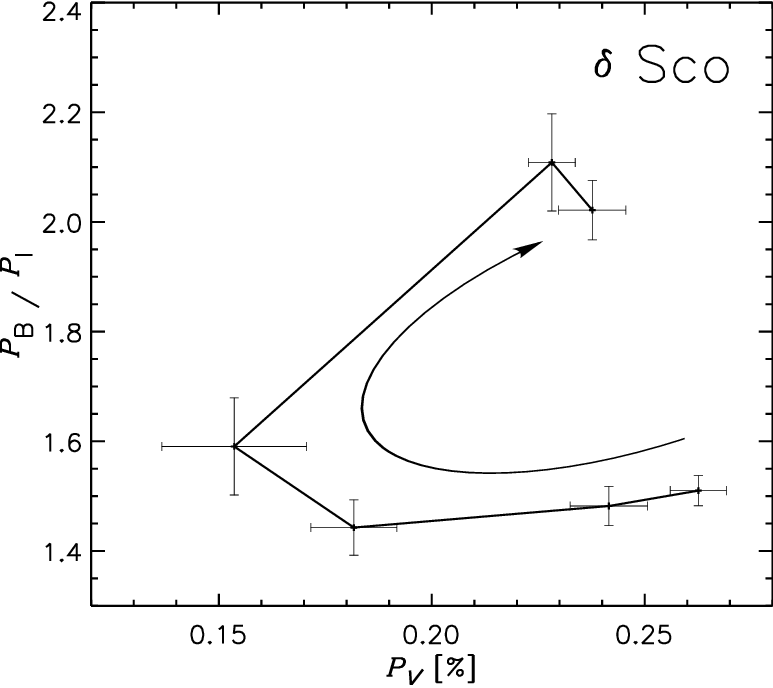}
 %\label{delsco}  
 \caption{Observed PCDs of $\pi$ Aqr (left panel) and  60 Cyg (middle panel), from \cite{Draper2013}, and PCD of  $\delta$ Sco (right panel) made from measurements obtained at the IAG polarimeter. For the three PCDs, arrows indicate the chronological succession of epochs. } 
 %The PCD presents a clockwise track starting at MJD=4975 (blue point). The other epochs in MJD are as follows: 5344,\rouge{?},5435,\rouge{?},5700.   4975, 5344, 5411, 5435, 5711, 5741
 \label{observational}
\end{center}
\end{figure*}

\subsection{$\pi$ Aquarii}

The PCD of $\pi$ Aqr  presents a very clear observational evidence of a partial loop obtained during a $\sim$180 day polarimetric flare (Fig.~\ref{observational}, left panel). From a pre-existent disk ($P_{V}=0.9\%$, $P_{\rm BD}=1.8$), the loop starts with a build-up phase until it reaches a maximum $P_{\rm BD}$ value of $2.2$.
After that, an irregular dissipation pattern (with small deviations probably due to short-lived mass injection events) is seen.
As already reported in \cite{2011ApJ...728L..40D} and in the present paper, the general clockwise loop is well explained by viscous processes. The exact PCD then depends on an interplay between the injected quantity of matter, the inclination angle, the $\alpha$ parameter and a good knowledge of the disk state before the outburst. Such in depth analysis will be carried out in the future. However, $\pi$ Aqr being a B1 type star \citep{1982ApJS...50...55S}, the range of the observed $P_{V}$ and $P_{\rm BD}$ values are consistent with the range of the values shown in Fig.~\ref{col-pol-loop3}. Since these latters  correspond to maximum values obtained for full building-up and dissipation cases, we conclude the $\pi$ Aqr polarimetric observations are compatible with our VDD predictions.
    
   % If we further assume an $\alpha$ of $1$ and an inclination angle close to 85$\degr$, a density of  $ \sim 5\times10^{-11}$$g.cm^{-3}$ would be a good estimation to repr.
      %  So one possible model able to reproduce $\pi$ Aqr's PCD would involve a typical and a high inclination angle.

  %84 day period binary, eccentricity probably very low , inclination angle to be between 50 and 75$\degr$ Bjorkman 2002
\subsection{60 Cygni}     
60 Cyg presents a more complex case (Fig.~\ref{observational}, middle panel). The first part of the PCD loop forms a clear clockwise loop that is well understood with a classical building-up/dissipation scenario, albeit with an irregular shape, suggestive of possible events of mass injection into the disk. In this part, the slopes are quite high for a  B1 spectral type star \citep{1982ApJS...50...55S} when one compares with Fig.\ref{col-pol-loop3}. We speculate this could be a signature of a high base density (i.e., high mass injection rates) and/or low inclination angles. The remaining data points follow a quite irregular track in the PCD, displaying the enigmatic behavior of an increase of $P_{\rm BD}$ with a simultaneous decrease of $P_{V}$. The results of \S~3.1 and 3.3 suggest possible ways to explain such a behavior (e.g., irregular mass injection rates), but the scarcity of the data points prevents any well-founded interpretation.

% The total duration is $\sim900$ days.

\subsection{$\delta$ Scorpii} 
$\delta$ Sco is a B0.2\,IVe star with an inclination angle of about 35$\degr$ \citep{2006ApJ...652.1617C}. A several-year-long campaign led at the Pico dos dias Observatory (LNA, Brazil) allowed to monitor the polarimetric activity of $\delta$ Sco in the $B$, $V$, $R$, $I$ filters \citep{bed2012}. At JD $\sim$ 2 445 500, the star was in a intermediary level of activity, having built a large disk in the course of the previous 10 years, when a photometric increase was observed in March 2010. Since the star is close to pole-on viewing, this increase is likely associated with a strong outburst (Paper I).

Polarimetric measurements were obtained before, during and after the march 2010 outburst (Fig.~\ref{observational}, right panel). In the PCD, the data shows the following chronological behavior:
\begin{itemize}
\item the $P_{V}$ level decreases after the photometric outburst had started while the $P_{B}/P_{I}$ level stays essentially the same, 
\item as the outburst proceeds, both $P_{B}/P_{I}$ and $P_{V}$ increases.
\end{itemize}

This behavior is qualitatively very similar to the outburst scenario described in Fig.~\ref{BDVD3}. When an outburst happens on top of an already existing disk (which is the case for the 2010 outburst of $\delta$ Sco), what is initially seen is a \emph{leftward} track in the PCD, followed by an up-right track, characteristic of disk build-up and subsequent dissipation. The range of $P_{V}$ levels are much lower than on Fig.~\ref{BDVD3} which could indicate that the $\delta$ Sco disk was not in a very dense state. Even though no attempt is made here to fit the data, it is worth noting that the outburst scenario, as suggested by photometry \citep{bed2012}, is a quite attractive explanation for the unusual track in the PCD seen for $\delta$ Sco, which, in turn, suggests that this track is consistent with a viscous disk scenario. This certainly deserves further scrutinization.

%However the timescales involved are different: the measurements of Fig.~\ref{observational} represent 2 years whereas the corresponding part of the PCD shown in Fig.~\ref{BDVD3} represent 0.25 year. This suggests a viscous $\alpha$ coefficient much lower than 1 for the $\delta$ Sco disk at these epochs.   (this is coherent with $\alpha <1 $  value and/or with a lower base density). 
%On the other hand, $P_{B}/P_{I}$ levels vary more on the $\delta$ Sco PCD than on Fig.~\ref{BDVD3}. This could be the signature of a strong outburst with a mass loss rate $\dot{M} > 3.3\times10^{-8} M_{\odot}$/year. We conclude by saying that even though we don't precisely reproduce the PCD shown in Fig.~\ref{observational}, these data are likely to represent the polarimetric counterpart of the outburst that started at MJD$\sim$5000.

\subsection{$\psi$ Persei}

\begin{figure}[h!]
%\centering
% \vspace*{-2.0 cm}
\begin{center}
 \includegraphics[height=1.7in]{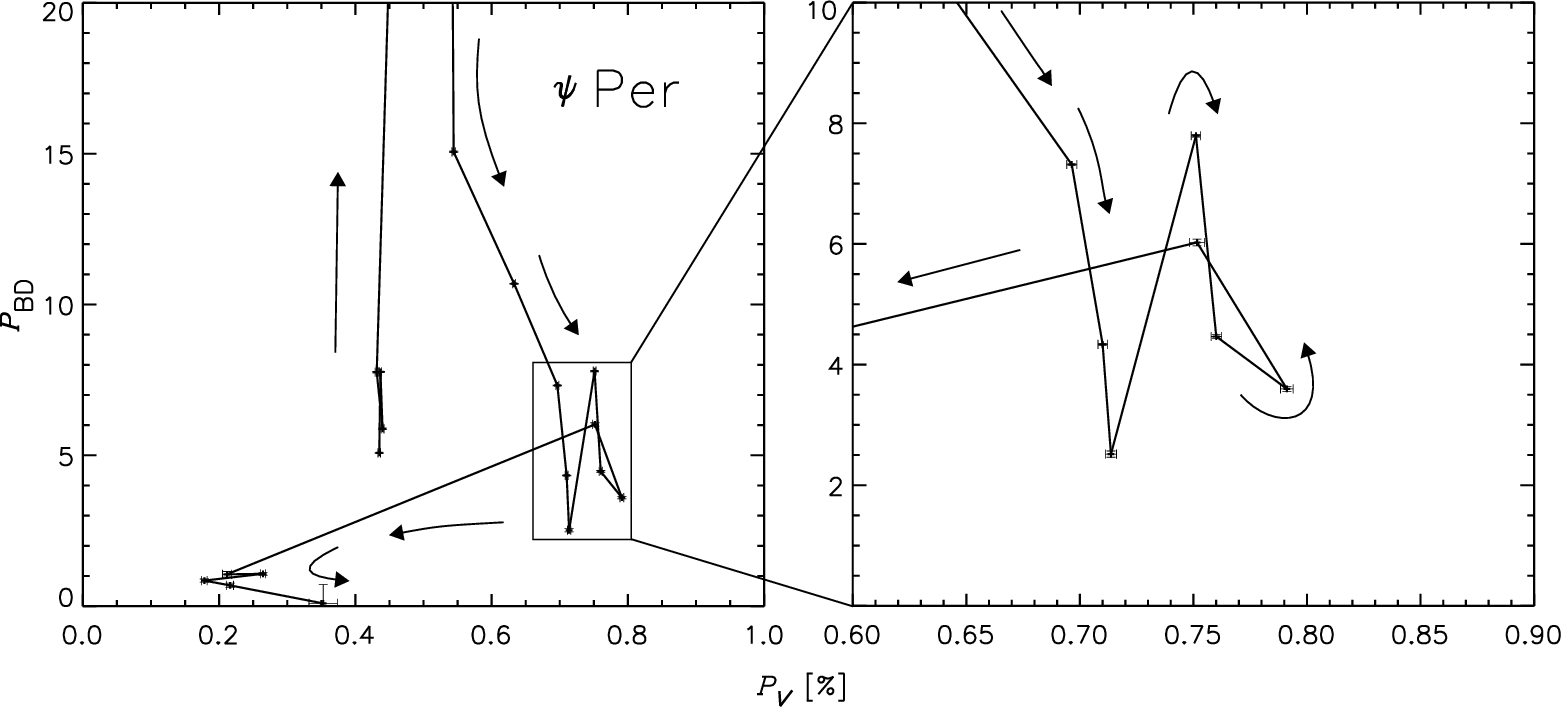} 
 \caption{PCD of $\psi$ Per built from HPOL observations that spanned 15 years.}
   \label{psiper}
\end{center}
\end{figure}

Another observed PCD that we can compare to our models, especially with Fig.~\ref{col-pol-loop4}, is  $\psi$ Per \citep[B5 star with a disk inclination angle of $75 \pm 8 \degr$,][]{2011A&A...529A..87D}. Figure~\ref{psiper} represents a 15 year long polarimetric monitoring of this star. The first characteristic of this PCD is the gigantic variation seen in $P_{\rm BD}$ while $P_{V}$ steadily increases over a period of 10 years, indicating that the disk is overall building up. The data point which is off scale corresponds to  $P_{\rm V} \sim 0.5$ and $P_{\rm BD} \sim 80$. Secondly, after $P_{\rm BD}$ decreased, a series of lower amplitude $P_{\rm BD}$ variations is seen (see zoom of Fig.~\ref{psiper}), until both $P_{\rm BD}$ and $P_{\rm V}$ decline on a few month time length. The track that $\psi$ Per follows on the PCD presents many similarities with the curve corresponding to the B5 spectral type on Fig.~\ref{col-pol-loop4}. Even though the time-scale and amplitude of the $P_{\rm BD}$ jump is not reproduced specifically by our model, the fact that a high $P_{\rm BD}$ variation is concomitant with a steady increase of  $P_{\rm V}$ supports the fact that $\psi$ Per was experiencing a building-up phase seen at a high inclination angle and with a very high disk density. We also mention that this peaky shape for a PCD is typical of late-type stars as shown in Fig.~\ref{col-pol-loop4}. Moreover, the zoomed panel of Fig.~\ref{psiper} also shows a pattern that is very similar to the counter-clockwise secondary loop in the PCD presented on Fig.~\ref{col-pol-loop4}. This type of behavior was only observed for models involving an inclination angle of $i$ $>$ 80 $\degr$ and a base density higher than $1\times10^{-10}$$\,\rm g\,cm^{-3}$. However, we speculate that this sort of PCDs could be observed for lower inclination angles if the disk is denser ($ \rho_0>  1\times10^{-10}$$\,\rm g\,cm^{-3}$).

%This conclusion is further reinforced by the fact that, when $\psi$ Per exhibited the $P_{\rm BD}$ jump, the bluewards part of the Balmer Discontinuity presented an almost null polarization. Based on the analytical approach detailed in \S~\ref{origin}.  we were able to see that same behavior for high inclination angles (90 $ \pm 5\degr$). 

It is important to mention that loop-shaped PCDs can also be qualitatively reproduced using spiral oscillation models as demonstrated in the Fig.~9 of \cite{2013ApJS..208....3H}. In this approach, a one armed density wave that is confined to a region extending out to 10 $R_\star$ generate distinct shapes for the PCDs as well as characteristic signatures in the polarization angle variation. We conclude by saying that a detailed modeling specific to each observed Be star is needed to understand the origin of the polarimetric variability. PCDs and polarization angle measurements are critical observable quantities to disentangle a scenario involving a variable mass injection rate in an axisymmetric disk (with no polarization angle variation expected) from a scenario including a one-armed oscillation.

\section{Conclusions}

In this paper, we first review the mechanisms at the origin of the continuum polarization levels observed for Be stars under the assumption of a steady-state VDD model. For different disk densities and spectral types, we detail how the electron scattering, bound-free and free-free opacities shape the polarized spectrum. Analyzing further these opacities, we describe how they vary radially to establish the regions of the disk that are electron-scattering dominated and the regions that are bound-free dominated. For low densities (typically $1 \times 10^{-12} \rm g\,cm^{-3}$), all models are electron scattering dominated and the polarization spectrum is roughly flat regardless the spectral type. 
For larger densities (typically greater than $1 \times 10^{-11} \rm g\,cm^{-3}$), the bound-free opacity becomes the dominant opacity in the disk, at least in its innermost part. Based on this opacity description, we can define two polarimetric features that depend on different disk properties: $P_V$, the polarization level in the visible (or any other wavelength), is a measure of the scattering mass of the disk; and $P_{\rm BD}$ (the polarization change across the Balmer jump), a measure of the \emph{ color} of the polarization spectrum, which depends on the relative importance of the bound-free opacity to the total opacity. 
With a more realistic modeling based on a coupling of hydrodynamics and radiative transfer simulations, we show how these polarimetric features evolve with time for different disk mass injection scenarios. Different mass injection histories (constant, periodic or episodic) result in specific behaviors of the polarimetric observables. This led us to introduce an extension of the concept of BJV diagrams \citep{2011ApJ...728L..40D} that we named PCD (Polarization Color Diagram) as an analogy to color magnitude diagrams. The PCD plots $P_{\rm BD}$ (or other measure of the polarization color) versus the polarization level. PCDs constitute a powerful diagnosis tool to derive different physical parameters of the disk such as the inclination angle, the viscous coefficient $\alpha$, the disk base density and the spectral type of the central star on top of the mass injection history. Typically, the polarimetric features of a star evolving from a diskless phase (normal B star) to a Be phase (with a disk present) and back to a B phase will appear as a loop in the PCD, the upper part of the loop being associated with the disk construction and the lower part with the dissipation. In some particular cases of very high densities and high inclination angles, morphologic variations of the PCDs with the apparition of a secondary loop were found. Albeit the useful diagnostic potential, the morphologies seen in the PCD bear some degeneracy with respect to some parameters. For this reason, we stress that the more observables and the better time coverage, the easier to infer a dynamical scenario and physical parameters of the system. 
%For instance, similar PCD loops can be obtained for a system with a rather low inclination angle ($\sim 30\degr$) and an early spectral type (B2) and for a system with a higher inclination  ($\sim 70\degr$) and a later spectral type (B5). 
The theoretical predictions were confronted with observed PCDs of four stars. While $\pi$~Aqr exhibits quite a clear building-up/dissipation pattern in its PCD,  60 Cyg presents a more irregular case that is likely to be the result of mass injection rate variations. We then analyse the PCD of  $\delta$ Sco and speculate that it probably represents the polarimetric counterpart of an outburst seen with photometry. Finally, we point out that the huge variations and the counter clockwise structure of $\psi$ Per's PCD can be explained with our predictions for a B5 spectral type star, with a high inclination angle and a high density. However, a detailed modeling remain to be carried out for each specific star  to achieve an accurate reproduction of the observed PCDs.

%Viscous diffusion is improving, a non isothermal will be soon released.
%Link toward the website : fix this color scale pbm wit IDL.
%Further scenarios to investigate : more complex mass loss decretion, blob  puffing.
%Opening on interferometry and spectroscopy.

\begin{acknowledgements}
    XH thanks FAPESP for supporting this work through the grants  2009/07477-1 and 2010/19029-0. BCM acknowledges support from CNPq (grant 133338/2012-6). ACC acknowledges support from CNPq (grant 307076/2012-1) and Fapesp (grant 2010/19029-0). DB acknowledges support from CNPq (grant 134761/2012-0). This work has made use of the computing facilities of the Laboratory of Astroinformatics (IAG/USP, NAT/Unicsul), whose purchase was made possible by the Brazilian agency FAPESP (grant 2009/54006-4) and the INCT-A.
     \end{acknowledgements}

\appendix

\renewcommand\thefigure{\thesection.\arabic{figure}}   
\setcounter{figure}{0}    

\section{The Radial Dependence of the Bound-Free Absorption Coefficients} \label{appendix}

 \begin{figure}[t]
% \vspace*{-2.0 cm}
\begin{center}
\includegraphics[width=3.5in]{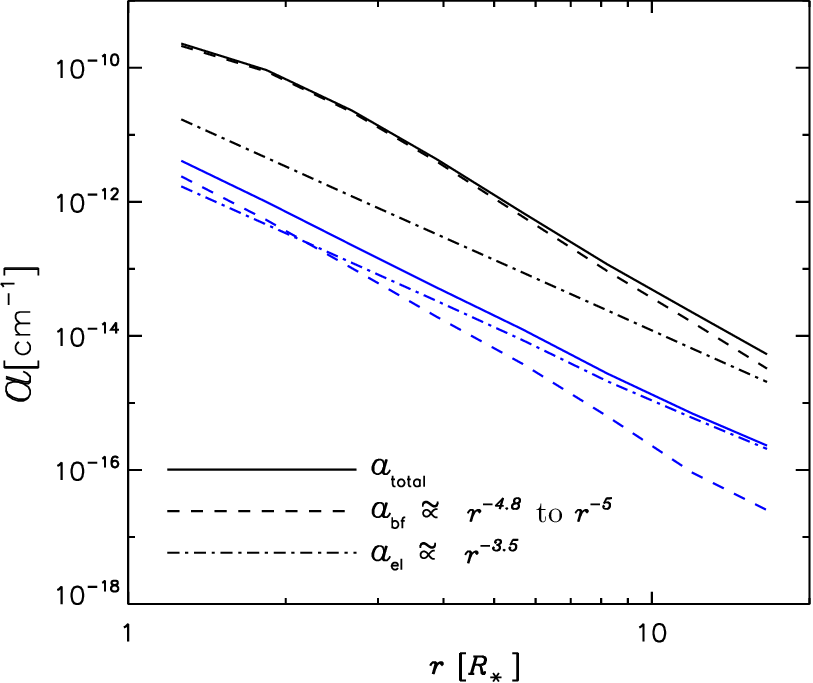}
\includegraphics[width=3.5in]{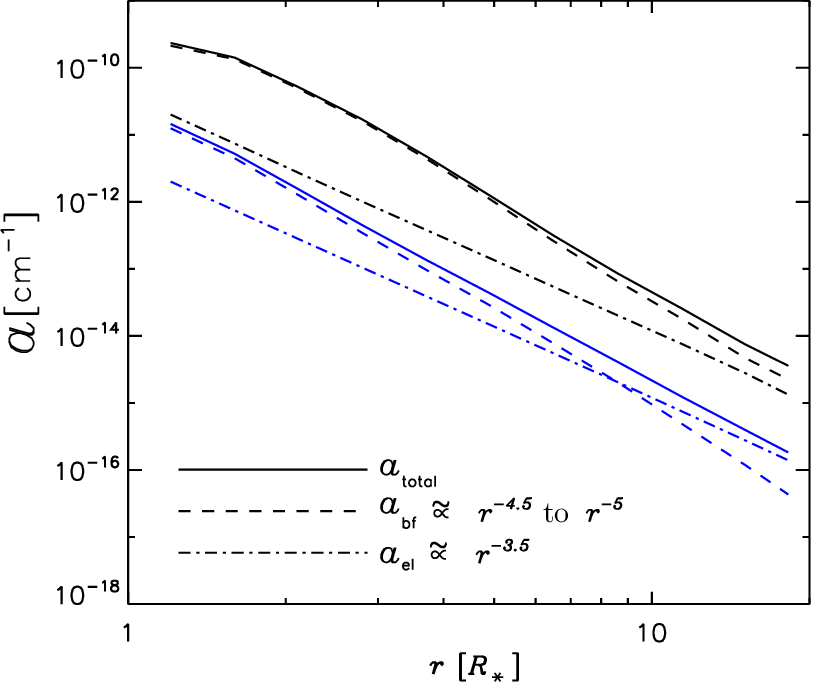}
 \caption{Same as Fig.~\ref{loci} for a B2 star (left) and a B3 star (right) and two different base densities: $\rho_{0} = 4.2 \times 10^{-11} \rm g\,cm^{-3}$ in black, and $4.2 \times 10^{-12} \rm g\,cm^{-3}$ in blue.}
\label{loci_dens}
\end{center}
\end{figure}

Figure~\ref{loci} showed that the radial dependence of the bound-free absorption coefficient is much steeper than the radial dependence of the density itself.
A similar figure (Fig.~\ref{loci_dens}) compares how the scattering and bound-free absorption coefficients changes with changing density for two spectral types. It is interesting to note that the bound-free absorption coefficient becomes dominant on a larger extent of the disk for both base densities for later spectral types.

The radial dependence of the bound-free absorption coefficients in these two figures, consider the expression for the bound-free absorption coefficient in the Paschen continuum ($\lambda_2 < \lambda < \lambda_3$, where $\lambda_2 = 3646\,\AA$ and $\lambda_3 = 8203\,\AA$ are the \ion{H}{1} $n=2$ and 3 photoionization thresholds).
Taking into account, for simplicity, only the contributions of \ion{H}{1} levels $n=2$ and 3 to the opacity \citep[e.g.,][Eq.~30]{1994ApJ...436..818B}

\begin{equation}
a_{\rm bf}(\lambda, r) = n(r) \left[ % \left\{ 
%\begin{array}{ll}
%N_3 a_3 \left(\frac{\nu_3}{\nu}\right)^3 & \nu_2>\nu>\nu_3 \\ 
N_2 b_2  \left(\frac{\lambda}{\lambda_2}\right)^3 + N_3  b_3  \left(\frac{\lambda}{\lambda_3}\right)^3  %& \nu>\nu_2 
\right] \,,
%\end{array} w
%\right.
\label{bfsct}
\end{equation}
where the photoionization cross sections are $b_2 = 1.4\times 10^{-17}\,\rm cm^{-2}$ and $b_3 = 2.2\times 10^{-17}\,\rm cm^{-2}$, $n(r)$ is the particle number density, and $N_2$ and $N_3$ are the fractional occupation numbers of \ion{H}{1}.

\begin{figure}[t]
% \vspace*{-2.0 cm}
\begin{center}
 \includegraphics[width=3.5in]{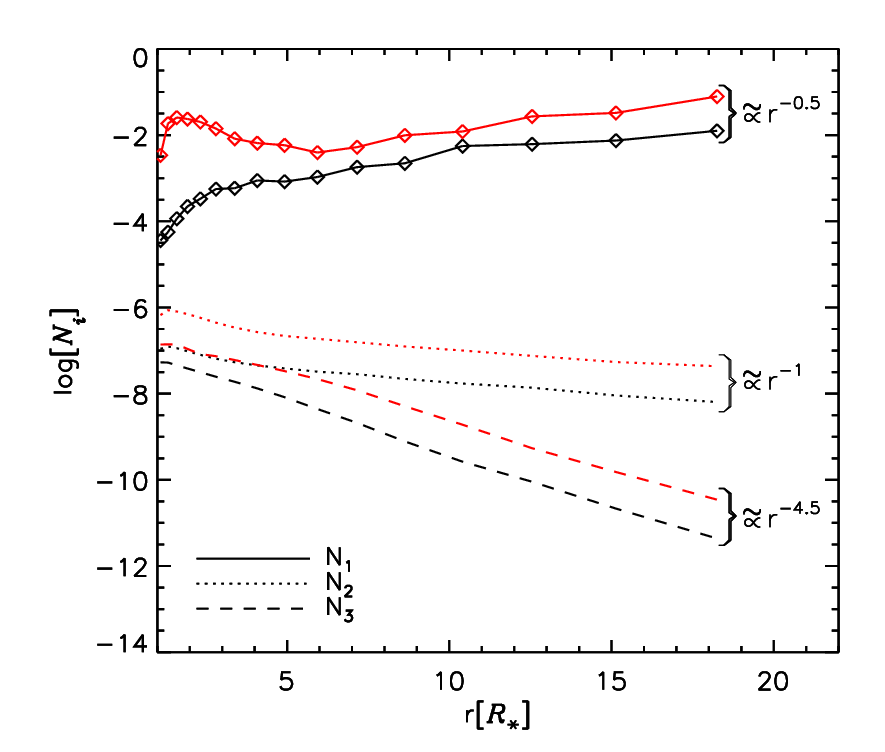} 
  \includegraphics[width=3.5in]{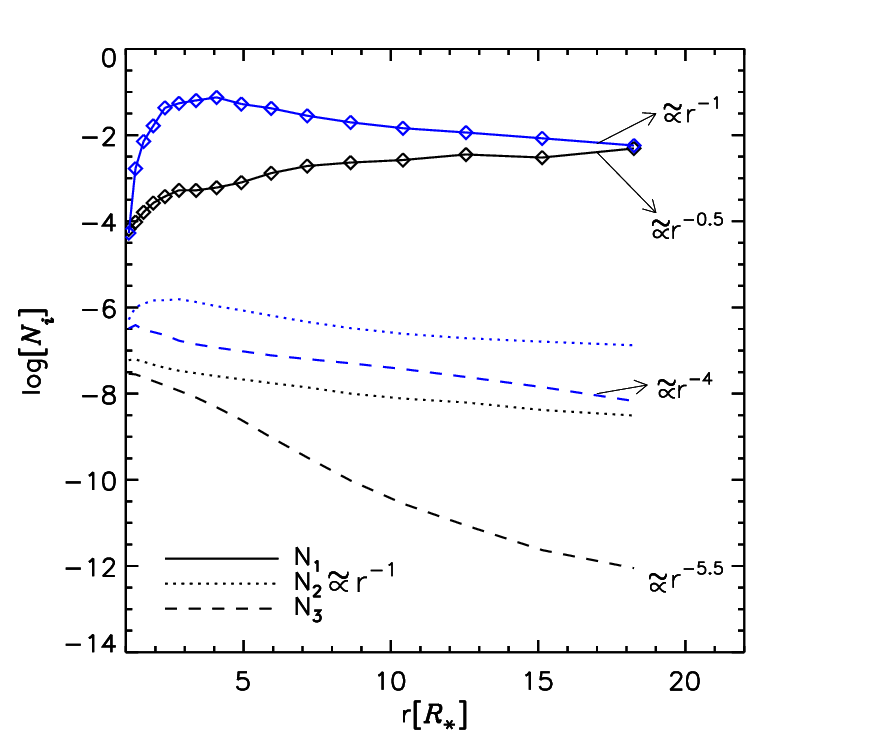}
 \caption{Hydrogen level populations (up to level 3) vs. distance from the star.
 %Upper panel: for a B2 star with a disk base density of $\rho_{0}$ = $8.4\times10^{-12} \rm g \,cm^{-3}$ and $\rho_{0}$ = $4.2\times10^{-11} \rm g \,cm^{-3}$.  \rouge{Version without the labels of numerical densities  for 5e13 (in blue) and 1e13 p/cm3 in red (like Fig.2)?} Lower panel: 
{\it Left: }
Results for different spectral types (B2 in black and B4 in red) for a steady-state VDD with $\rho_{0}=8.4\times10^{-12} \rm g \,cm^{-3}$. 
{\it Right: }
Results for different base densities ($\rho_{0} = 4.2 \times 10^{-12} \rm g \,cm^{-3}$, black, and $8.4\times10^{-11} \rm g \,cm^{-3}$, blue) for a B2 star. Power-law indexes are indicated.}
\label{popu}
\end{center}
\end{figure}

The \ion{H}{1} occupation numbers of the same models shown in Fig.~\ref{loci} and in the left panel of Fig.~\ref{loci_dens} are plotted in Fig.~\ref{popu}.
$N_3$ is much smaller than $N_2$ everywhere in the disk, so, to a first approximation, the contribution of the atoms in this level (and above) can be ignored.
Thus, the radial dependence of $a_{\rm bf}$ is thus controlled by the (explicit) radial dependence of $n(r)$ and the (implicit) radial dependence of $N_2$
\begin{equation}
a_{\rm bf}(\lambda, r) \propto n(r)N_2 \propto r^{-(A+B)}\,,
\end{equation}
where $A$ is the slope of the density distribution and $B$ is the radial variation of $N_2$.
For a steady-state disk, $A=3.5$ (\S~\ref{origin}). The value of $B$ is not constant, as is apparent from the fact that the $N_2$ curves in Fig.~\ref{popu} are not straight lines. However, representative values of $B$ can be found by fitting the $N_2$ with a power-law. The results, shown in Fig.~\ref{popu} gives $B \sim 1$. Thus, $a_{\rm bf}$ falls approximately as $r^{-4.5}$.

%% in parentheses, after the keyword, but they will not be verified.



%\appendix

%\section{Appendix material}

\end{document}